\documentclass[a4paper,11pt]{article}
\pdfoutput=1 

\usepackage{jheppub} 

\usepackage[T1]{fontenc} 

\usepackage{subfigure}

\title{\boldmath Relative Entanglement Entropies in 1+1-dimensional conformal field theories}
\author{Paola Ruggiero and Pasquale Calabrese}

\affiliation{International School for Advanced Studies (SISSA) and INFN,\\Via Bonomea 265, 34136, Trieste, Italy}

\emailAdd{paola.ruggiero@sissa.it}
\emailAdd{calabrese@sissa.it}

\abstract{
We study the relative entanglement entropies of one interval between excited states of a 1+1 dimensional 
conformal field theory (CFT). 
To compute the relative entropy $S(\rho_1 \| \rho_0)$ between two given reduced density matrices  $\rho_1$ and $\rho_0$ 
of a quantum field theory, we employ the replica trick which relies on the path integral representation of 
${\rm Tr} ( \rho_1 \rho_0^{n-1} )$ and define a set of R\'enyi relative entropies $S_n(\rho_1 \| \rho_0)$. 
We compute these quantities for integer values of the parameter $n$ and derive via the replica limit, the 
relative entropy between excited states generated by primary fields of a free massless bosonic field. 
In particular, we provide the relative entanglement entropy of the state described by the primary operator $i \partial\phi$, 
both with respect to the ground state and to the state generated by chiral vertex operators.
These predictions are tested against exact numerical calculations in the  XX spin-chain finding perfect agreement.
}

\begin{document} 
\maketitle
\flushbottom

\section{Introduction}

In the last years the irruption in other research fields of concepts and methods coming from quantum information turned out to be 
very fruitful. So far, particular attention has been devoted to the characterization of different measures of 
entanglement \cite{plenio-2007} (among them especially to the entanglement entropy) in physical states of extended 
systems such as quantum field theories and many-body quantum matter.

For example, in the condensed matter community (see e.g. \cite{amico-2008, calabrese-2009, eisert-2010,rev-lafl} as reviews) 
entanglement has been largely employed as a tool for detecting quantum phase transitions and to deduce information about 
the underlying conformal field theory (CFT), by looking at the universal behavior of the entanglement 
entropy \cite{hlw-94,vlrk-03,cc-04,cc-09}; the study of the entanglement spectrum proved to give a deeper understanding of 
topological features of some condensed matter systems such as quantum Hall states \cite{lihaldane}; 
in out of equilibrium situations a deep connection emerged between entanglement and entropy production \cite{cc-05,kauf,ac-16}; 
other entanglement measures, such as entanglement negativity, allowed to deal with systems in mixed quantum states 
as well \cite{CCT}. 

Also in the high-energy/gravity community entanglement has found a wide variety of applications, particularly in connection to the 
black hole physics, where it is largely believed that it plays a fundamental role in the interpretation of the Bekenstein-Hawking 
entropy \cite{sorkin-1983, bombelli, sredniki} and in the AdS-CFT correspondence, where the Ryu-Takayanagi formula 
is still  one of the major result \cite{takayanagi}.
 
So far, the large majority of these studies focused  on the entanglement of a subsystem of a given quantum state.
It is a very natural question whether, more generally speaking, exploring other quantum information concepts 
could provide more insights when considering two different quantum states (obviously defined on the same Hilbert space).
In this respect, an interesting quantity to look at is the so-called \emph{relative entropy} \cite{relative-entropy}
that for two given (reduced) density matrices $\rho_1$ and $ \rho_0$, is defined as
\begin{equation}
S(\rho_1 || \rho_0) = {\rm Tr} (\rho_1 \log \rho_1) -{\rm  Tr }(\rho_1 \log \rho_0),
\end{equation}
which can be interpreted as a measure of distinguishability of quantum states, being a sort of (asymmetric) ``distance'' 
between $\rho_1$ and $\rho_0$.
It is not an entanglement measure itself, but nonetheless has connection with several entanglement measures \cite{vedral-2002,ae-05}.

The relative entropy attracted only recently the interests of the field theory community, but it is already taking a central role given 
the number of papers devoted to it, see e.g. \cite{casini-2016,clt-16,black-hole-thermodynamics,bekenstein-bound,hol-rel-entropy,lashkari2014,lashkari2016,ugajin2016,ugajin-higherdim,ugajin2016-2,abch-16, balasubramanian-14, caputa-16}. 
One of its advantages is that, contrarily to the entanglement entropy which in a quantum field theory framework suffers from the problem of ultraviolet divergences, the relative entropy is finite and therefore well defined also in field theory. 

The relative entropy is also related to the entanglement (or modular) Hamiltonian, or better to its variation between 
two quantum states. Indeed, it straightforwardly holds 
\begin{equation}
S(\rho_1 || \rho_0)= \Delta \langle{\cal H}_0\rangle-\Delta S\,,
\end{equation}
where $\Delta S\equiv S(\rho_1)-S(\rho_0)$ is the difference of von Neumann entropies $S(\rho)\equiv -{\rm Tr} \rho\log\rho$ and
$\Delta \langle{\cal H}_0\rangle$ is the variation of the modular Hamiltonian ${\cal H}_0$ (implicitly defined as 
$\rho_0=e^{-{\cal H}_0}/{{\rm Tr} e^{-{\cal H}_0}}$) relative to $\rho_0$, i.e., 
\begin{eqnarray}
 \Delta \langle{\cal H}_0\rangle&=& {\rm Tr} [(\rho_1-\rho_0){\cal H}_0]. 
\end{eqnarray}
This relation between relative entropy and modular Hamiltonian is the starting point 
of the recent (alternative) proofs of the Zamolodchikov's c-theorem \cite{c-theorem} in Ref. \cite{casini-2016}
and of the boundary g-theorem \cite{al-91} in Ref. \cite{clt-16}.
Furthermore, being the entanglement Hamiltonian a central object in many problems, as e.g. 
in Refs. \cite{lihaldane,bw-76,chm-11,wkpv-13,ct-16}, the knowledge of the relative entropy can help also in these circumstances.

The relative entropy may give useful insights also in the study of condensed matter systems. For example singularities in other measures of distinguishability among quantum states (as it is the case for the quantum fidelity \cite{fidelity}) have already been proposed as a signature of a quantum phase transition.

The relative entropy has also been considered in connection to the laws of black hole thermodynamics \cite{black-hole-thermodynamics} and the Bekenstein bound \cite{bekenstein-bound}, which can both be shown to follow from the properties of positivity and monotonicity of the relative entropy. Its holographic version has been discussed as well \cite{hol-rel-entropy}.

In a quantum field theory, the relative entropy can be obtained by a variation of the replica trick for the entanglement entropy \cite{cc-04}
which has been introduced by Lashkari \cite{lashkari2014} and later refined by the same author \cite{lashkari2016}. 
The main idea is to introduce the new quantity ${\rm Tr} (\rho_1 \rho_0 ^{n-1} )$ that for $n$ integer 
is a generalized partition function or correlation function on a $n$-sheeted Riemann surface which breaks the $Z_n$ 
symmetry among replicas. 
The relative entropy is given by the following replica limit \cite{lashkari2016}
\begin{equation} \label{rep_Srel}
S (\rho_1 \| \rho_0) = \lim_{n \to 1} \frac{1}{1-n} \log \frac{{\rm Tr} (\rho_1 \rho_0^{n-1}) }{{\rm Tr} (\rho_1^n)} = 
\lim_{n \to 1} - \frac{\partial}{\partial n} \frac{ {\rm Tr} (\rho_1 \rho_0^{n-1}) }{{\rm Tr} (\rho_1^n)} ,
\end{equation}
whenever the analytic continuation of the parameter $n$ from integer to complex values is obtainable.
This method is completely general and permits (at least in principle) the computation of the relative entropy in a 
generic quantum field theory.
However, up to now, only few direct calculations of relative entropy have been performed in 1+1 dimensional 
CFT \cite{lashkari2014, lashkari2016,ugajin2016,ugajin2016-2} and only very recently some results for arbitrary dimensions 
appeared \cite{ugajin-higherdim}.

In analogy to the entanglement R\'enyi entropies 
$$S_n(\rho)\equiv \frac{1}{1-n} \log {\rm Tr} (\rho^n),$$
we can define {\it R\'enyi relative entropies} as 
\begin{equation}
S_n (\rho_1 \| \rho_0) \equiv \frac{1}{1-n} \log \frac{{\rm Tr} (\rho_1 \rho_0^{n-1}) }{{\rm Tr} (\rho_1^n)}\,.
\label{RRE}
\end{equation}
While it is still unknown whether these quantities have a quantum information interpretation, they surely have two interesting 
features: 
i) when $\rho_0$ equals the identity $S_n$ reduces to minus the R\'enyi entropy of $\rho_1$, i.e.
$S_n (\rho_1 \| \rho_0={\mathbb I})=-S_n(\rho_1)$,  alike $S (\rho_1 \| \rho_0={\mathbb I})=-S(\rho_1)$;
(ii) its limit for $n\to1$ is $S (\rho_1 \| \rho_0)$.
The main drawback of $S_n (\rho_1 \| \rho_0)$ is that, contrarily to $S (\rho_1 \| \rho_0)$, is not always a positive function
(as we shall see in the following). 
This is similar to standard R\'enyi entropies that satisfy strong subadditivity \cite{lr-73} only for $n=1$.

In this paper, we perform a systematic study of the relative entanglement entropies and their R\'enyi counterpart between 
excited states  associated to primary operators in the free massless bosonic field theory in 1+1 dimensions,  
generalizing the analysis of previous works \cite{lashkari2014,lashkari2016,ugajin2016,ugajin2016-2}. 
Furthermore we provide the first explicit checks of the CFT results in concrete lattice models. 

The paper is organised as follows. 
In Section \ref{section:CFTapproach} we review the  CFT approach to the R\'enyi 
relative entropies between the reduced density matrices of two excited states associated to primary fields. 
In Section \ref{section:Examples} we present explicit calculations of relative entropy in the massless bosonic theory
and in particular for the derivative operator $i\partial \phi$. 
These CFT results are tested in Section \ref{section:comparison} against   
exact numerical calculations in the XX spin-chain, whose continuum limit is a free massless boson. 
Finally, we conclude and discuss some future perspectives in Section \ref{section:conclusions}.

\section{CFT approach to the relative entropy}
\label{section:CFTapproach}

We consider a one-dimensional system and a bipartition into two complementary regions $A$ and $\bar{A}$, inducing a 
bipartition of the Hlibert space as 
\begin{equation}
\mathcal{H} = \mathcal{H}_{A} \otimes \mathcal{H}_{\bar{A}}.
\end{equation}
Given  two generic states $| \psi_1 \rangle, | \psi_0 \rangle\in\mathcal{H}$, the  reduced density matrices (RDM)  of the subsystem $A$ 
are given respectively  by 
\begin{equation} \label{phipsi}
\rho_1= {\rm Tr}_{\bar{A}} | \psi_1 \rangle \langle \psi_1|,  \qquad\rho_0 = {\rm Tr}_{\bar{A}} | \psi_0 \rangle \langle \psi_0 |.
\end{equation}

We are interested in computing the relative entropies between two eigenstates of the CFT $|\psi_1\rangle$ and $|\psi_0\rangle$ using the 
replica approach \eqref{rep_Srel}.
To this aim we need  a path integral representation of ${\rm Tr} \left( \rho_1 \rho_0^{n-1} \right) $ for $ n \in \mathbb{N}$ 
(or more in general of ${\rm Tr} \left( \rho_1^m \rho_0^n \right)$, with $m, n \in \mathbb{N}$)
which is a generalization of that in Refs. \cite{abs-11,sierra2012} for R\'enyi entropies of excited states in CFT.
We are now going to review the main steps to construct ${\rm Tr} \left( \rho_1 \rho_0^{n-1} \right)$, 
closely following Refs. \cite{sierra2012,lashkari2016}.

Let us consider a 1+1 dimensional CFT in imaginary time $\tau$. As usual, we parametrize the two dimensional geometry by a 
complex coordinate $w=x+ i \tau$, where the domain of the spatial coordinate $x$ can be finite, semi-infinite or infinite.
The ground-state density matrix may be written as the path integral on the imaginary time as  \cite{cc-04,cc-09}
\begin{equation}
\langle \phi |\rho (\beta= \infty)| \phi'  \rangle = \frac{1}{Z} \int_{\varphi (- i \infty)= \phi}^{\varphi (i \infty) = \phi '} \mathcal{D} \varphi  e^{- S(\varphi)},
\end{equation}
with the value of the field fixed at $w= \pm i \infty$. 
$S (\varphi)$ is the euclidean action and $Z$ the normalization to have  ${\rm Tr} \rho (\beta)=1$.
This is nothing but the $\beta\to\infty$ limit of the thermal density matrix.

We will be interested only in excited states of the CFT  which are obtained 
by acting on the ground state with a generic primary operator $\Upsilon$ (i.e. $|\Upsilon\rangle\equiv \Upsilon(-i\infty)|0\rangle$), 
whose corresponding density matrix is 
\begin{equation} \label{rho_ex}
\langle \phi|  \rho_{\Upsilon} | \phi' \rangle = \langle \phi  |\Upsilon \rangle \langle \Upsilon| \phi' \rangle = 
\frac{1}{Z} \int_{\varphi (- i \infty)= \phi}^{\varphi (i \infty) = \phi '} \mathcal{D} \varphi \; \Upsilon ( i\infty)  \Upsilon^{\dagger} (-i  \infty) e^{- S(\varphi)}.
\end{equation}
As usual \cite{cc-04}, the RDM $\rho_{\Upsilon} (A)$ relative to the subsystem $A$ is given by closing cyclically  
$\rho_{\Upsilon}$ along  $\overline A$ and leaving an open cut along $A$.
Then ${\rm Tr} \rho^n_{\Upsilon} (A)$ is obtained by making $n$ copies of the RDM $\rho_{\Upsilon} (A)$ and sewing them 
cyclically along $A$. 
Following this standard procedure, we end up in a world-sheet which is a $n$-sheeted Riemann surface ${\mathcal{R}_n}$, 
and the desired moment of $\rho_{\Upsilon}(A)$ is \cite{sierra2012}
\begin{equation}
{\rm Tr} \rho_{\Upsilon}^n (A) \propto Z_n (A) \langle \prod_{k=1}^n \Upsilon (w_k) \Upsilon_k^{\dagger} (w'_k) \rangle_{\mathcal{R}_n},
\end{equation}
where the expectation value $\langle \cdots \rangle_{\mathcal{R}_n}$ is on the Riemann surface ${\mathcal{R}_n}$, 
$Z_n (A) \equiv \langle \mathbb{I} \rangle_{\mathcal{R}_n} $ (i.e. the $n$-th moment of the RDM of the ground state) 
and $w_k= i \infty, w'_k = - i \infty$ are points where the operators are inserted in the $k$-th copy.
Taking properly into account the normalization, this is
\begin{equation} 
{\rm Tr} \rho_{\Upsilon}^n (A) = \frac{ Z_n (A)}{Z_1^n} \frac{\langle \prod_{k=1}^n \Upsilon (w_k) \Upsilon_k^{\dagger} (w'_k) \rangle_{\mathcal{R}_n}}{\langle\Upsilon (w_1) \Upsilon^{\dagger} (w'_1) \rangle_{\mathcal{R}_1}^n}.
\end{equation}

Finally, it is convenient to consider the universal ratio between the moment of the RDM in the excited 
state $\Upsilon$ and the one of the ground state, i.e. 
\begin{equation} \label{F_Upsilon}
F_{\Upsilon}^{(n)} (A) \equiv \frac{{\rm Tr} \rho_{\Upsilon}^n}{{\rm Tr} \rho_{\mathbb{I}}^n} =
\frac{\langle \prod_{k=1}^n \Upsilon (w_k) \Upsilon_k^{\dagger} (w'_k) \rangle_{\mathcal{R}_n}}{\langle\Upsilon (w_1) \Upsilon^{\dagger} (w'_1) \rangle_{\mathcal{R}_1}^n},
\end{equation}
in which  the factors coming from the partition functions cancel out.

In order to calculate the correlators appearing in \eqref{F_Upsilon} in the case of $A$ being a single interval  
$A= \{ x \in (u, v) \}$, one considers the following sequence of conformal maps
\begin{equation}
w = x + i t \to \zeta = \frac{\sin[ \pi(w- u)/L]}{\sin [ \pi(w- v)/L] } \to z = \zeta^{1/n},
\label{maps}
\end{equation}
where $\zeta (w)$ brings $(u, v) \to (- \infty, 0)$ and $z(\zeta)$ is a \emph{uniformizing mapping} which maps the $n$-sheeted Riemann surface into the complex plane. 
According to these maps
\begin{equation}
\begin{cases}
w_{k} =  i \infty \to z_{k, n} = e^{i \frac{\pi}{n} (x+ 2(k-1))} \\
w'_{k} = - i \infty \to z_{k, n} = e^{i \frac{\pi}{n} (-x+ 2(k-1))}
\end{cases}
\; k = 1, \cdots, n, \quad x= \frac{v-u}{L} \equiv \frac{\ell}{L} .
\end{equation}
We shall use the transformation properties of the primary fields under conformal maps
\begin{equation}
\Upsilon (w, \bar{w}) = \left(  \frac{dz}{dw} \right)^{h} \left(  \frac{d\bar{z}}{d\bar{w} } \right)^{\bar{h}} \Upsilon (z, \bar{z}),
\end{equation}
being $(h, \bar{h})$ the scaling dimensions of $\Upsilon$.
In our case this becomes \cite{sierra2012}
\begin{equation}
\Upsilon (w_k, \bar{w}_k)  = \Big(  \frac{z_{k, n}}{n}  \Lambda \Big)^h \Big(  \frac{ \bar{z}_{k, n}}{n} \bar{ \Lambda} \Big)^{\bar{h}}
 \Upsilon ( z_{k, n}, \bar{z}_{k, n} ),
\end{equation}
with
\begin{equation}
\Lambda = \frac{4 \pi}{L} \sin(\pi x) e^{- 2 \pi |w|/L} e^{i \pi (u+ v)/L} .
\end{equation}

Finally the complex plane can be mapped to a cylinder of circumference $2 \pi$ by $t = - i \ln z$
which implies
\begin{equation}
\Upsilon (t, \bar{t}) = e^{i \pi (h - \bar{h})} z^{h } \bar{z}^{\bar{h}} \Upsilon (z, \bar{z}).
\end{equation} 
Combining all the above transformations, for our geometry of an interval $A$ of length $\ell$ embedded in a finite 
system of length $L$, we end up in  \cite{sierra2012}
\begin{equation} \label{final_excited}
F_{\Upsilon}^{(n)} (x) = n^{- 2n (h + \bar{h})} 
 \frac{\langle \prod_{k=1}^n \Upsilon (t_{k, n}) \Upsilon^{\dagger} (t'_{k,n}) \rangle_{\text{cyl}}}{\langle\Upsilon (t_{1, 1}) \Upsilon(t'_{1, 1}) \rangle_{\text{cyl}}^n},
\end{equation}
where we recall $x=\ell/L$ and
\begin{equation}
t_{k, n} = \frac{\pi}{n} (x+ 2(k-1)), \quad t'_{k, n} =  \frac{\pi}{n} (- x+ 2(k-1)), \quad k= 1, \dots , n.
\label{tdef}
\end{equation}
The above result has been generalized in the literature to many other circumstances such as generic states 
generated also by descendant fields \cite{p-14,p-16}, boundary theories \cite{txas-13,top-16}, 
and systems with disorder \cite{rrs-14}.

We now turn to the path integral representation of ${\rm Tr}\left( \rho_1^m \rho_0^n \right)$, which is a simple generalization of  
${\rm Tr} \rho^n$ discussed above. 
In this case, in fact, instead of $n$ copies of the RDM $\rho_0$ only, one considers further $m$ copies of $\rho_1$ and 
joins them cyclically as before.
Considering two CFT excited states of the form \eqref{rho_ex} obtained from the action of two primaries $\Upsilon_0$ 
and $\Upsilon_1$ , the final result is a path integral on a Riemann surface with $(m+n)$ sheets with the insertion 
of $\Upsilon_1, \Upsilon_1^{\dagger}$ 
on $m$ sheets and $\Upsilon_0 , \Upsilon_0 ^{\dagger}$ on the remaining $n$ sheets, i.e. \cite{lashkari2016}
\begin{equation}
{\rm Tr} \left(  \rho_1^m \rho_0^{n} \right) \propto Z_n (A) \Big\langle  \prod_{k=1}^m \Upsilon_1 (w_k) \Upsilon_1^{\dagger} (w'_k)   
\prod_{i=1+m}^{n+m} \Upsilon_0  ( w_i ) \Upsilon_0 ^{\dagger} (w_i) \Big\rangle_{\mathcal{R}_n}.
\end{equation}
Keeping track of the normalization we get
\begin{equation}
{\rm Tr} \left(  \rho_1^m \rho_0^{n} \right) =
\frac{Z_n (A)}{Z_1^{m+n}}
\frac{ \langle \prod_{k=1}^{m} \Upsilon_1(w_k) \Upsilon_1^{\dagger} (w'_k) \prod_{i= 1+m}^{n+m} \Upsilon_0  (w_i) \Upsilon_0 ^{\dagger} (w'_i) \rangle_{\mathcal{R}_n}  }{
\langle \Upsilon_1 (w_1) \Upsilon_1^{\dagger} (w'_1) \rangle_{\mathcal{R}_1}^{m} \langle \Upsilon_0  (w_1) \Upsilon_0 ^{\dagger} (w'_1) \rangle_{\mathcal{R}_1}^{n} }.
\end{equation}
In particular, for the (R\'enyi) relative entropy between $\rho_1$ and $\rho_0$, we compute the universal ratio 
\begin{equation} \label{2ndreplicaSrel}
G^{(n)} ( \rho_1 \| \rho_0 ) \equiv \frac{{\rm Tr} \left(  \rho_1 \rho_0^{n-1} \right) }{{\rm Tr} \left( \rho_1 ^n \right)} = 
\frac{ \langle \Upsilon_1(w_1) \Upsilon_1^{\dagger} (w'_1) \prod_{i= 2}^{n} \Upsilon_0  (w_i) \Upsilon_0 ^{\dagger} (w'_i) 
\rangle_{\mathcal{R}_n} 
\langle \Upsilon_1 (w_1) \Upsilon_1^{\dagger} (w'_1) \rangle_{\mathcal{R}_1}^{n-1}  }{
\langle \prod_{i=1}^n  \Upsilon_1 (w_i ) \Upsilon_1^{\dagger} (w'_i) \rangle_{\mathcal{R}_n} \langle \Upsilon_0  (w_1) \Upsilon_0 ^{\dagger} (w'_1) \rangle_{\mathcal{R}_1}^{n-1} }.
\end{equation}
 
Also in this case, to compute \eqref{2ndreplicaSrel}, we use the conformal maps $w \to z \to t$ \eqref{maps}, which bring it to the final form
\begin{multline} 
\label{2ndreplicaSrel-t}
G^{(n)} ( \rho_1 \| \rho_0 ) =
 n^{2(n-1) ( (h_{1} + \bar{h}_{1} ) - (h_{0 } + \bar{h}_{0 } )) }\\
\frac{ \langle \Upsilon_1(t_{1,n}) \Upsilon_1^{\dagger} (t'_{1, n}) \prod_{i= 2}^{n} \Upsilon_0  (t_{i, n}) \Upsilon_0 ^{\dagger} (t'_{i, n}) \rangle_{\text{cyl} } \langle \Upsilon_1(t_{i, n}) \Upsilon_1^{\dagger} (t'_{i, n}) \rangle_{ \text{cyl}  }^{n-1}  }{
\langle \prod_{i=1}^n  \Upsilon_1 (t_{i, n}) \Upsilon_1^{\dagger} (t'_{i, n}) \rangle_{\text{cyl} } \langle \Upsilon_0  (t_{i, n}) \Upsilon_0 ^{\dagger} (t'_{i, n}) \rangle_{\text{cyl} }^{n-1} },
\end{multline}
being $h_{1}$ and $h_{0 }$ the scaling dimensions of $\Upsilon_1$ and $\Upsilon_0 $ respectively.
Note that $G^{(1)} ( \rho_1 \| \rho_0 )=1$ for any $\Upsilon_{0,1}$, as it should.

As already mentioned, the relative entropy is not symmetric in $\rho_1$ and $\rho_0$.
Therefore we are going to consider the two (generically different) quantities $S (\rho_1 \| \rho_0)$ and $S (\rho_0 \| \rho_1)$, 
obtained via replica limit from $G^{(n)} ( \rho_1 \| \rho_0 )$ and $G^{(n)} (\rho_0 \| \rho_1 )$  respectively.
Notice that the universal ratio $G^{(n)} ( \rho_1 \| \rho_0 )$ gives the R\'enyi relative entropy \eqref{RRE} 
as 
\begin{equation}
S_n( \rho_1 \| \rho_0 )= \frac1{1-n}\log G^{(n)} ( \rho_1 \| \rho_0 )\,.
\end{equation}  

In the limiting case when one of the states, say $\rho_1$, is the ground state, these universal ratios simplify as follows
\begin{eqnarray}
G^{(n)} ( \rho_{GS} \| \rho_0 )  &=& \frac{ \langle \prod_{i=1}^{n-1} \Upsilon_0  (w_i) \Upsilon_0 ^{\dagger} (w'_i) \rangle_{\mathcal{R}_n} }{\langle \Upsilon_0  (w_1) \Upsilon_0 ^{\dagger} (w'_1) \rangle_{\mathcal{R}_1}^{n-1} }, \\
G^{(n)} ( \rho_0 \| \rho_{GS} )  &=& \frac{\langle \Upsilon_0  (w_1) \Upsilon_0 ^{\dagger} (w'_1) \rangle_{\mathcal{R}_n} \langle \Upsilon_0  (w_1) \Upsilon_0 ^{\dagger} (w'_1) \rangle_{\mathcal{R}_1}^{n-1}  }{ \langle \prod_{i=1}^{n} \Upsilon_0  (w_i) \Upsilon_0 ^{\dagger} (w'_i)  \rangle_{\mathcal{R}_n} },
\end{eqnarray}
which after the usual mappings become
\begin{eqnarray} \label{rhosigmaRel2}
 G^{(n)} ( \rho_{GS} \| \rho_0 ) &=& n^{- 2 (n-1) (h_{0 }+ \bar{h}_{0 })}  
 \frac{ \langle \prod_{k=1}^{n-1} \Upsilon_0  (t_{k, n}) \Upsilon_0 ^{\dagger} (t'_{k, n}) \rangle_{\text{cyl}} }{\langle \Upsilon_0  (t_{1, 1}) \Upsilon_0 ^{\dagger} (t'_{1, 1}) \rangle_{\text{cyl}}^{n-1} }, \\
G^{(n)} (\rho_0 \| \rho_{GS} ) &=& n^{2 (n-1) (h_{0 } + \bar{h}_{0 }) } 
\frac{\langle \Upsilon_0  (t_{1,n}) \Upsilon_0 ^{\dagger} (t'_{1,n}) \rangle_{\text{cyl}} \langle \Upsilon_0  (t_{1,1}) \Upsilon_0 ^{\dagger} (t'_{1,1}) \rangle_{\rm cyl  }}{ \langle \prod_{k=1}^{n} \Upsilon_0  (t_{k,n}) \Upsilon_0 ^{\dagger} (t'_{k, n})  \rangle_{\rm cyl }}. \label{sigmarhoRel2}
 \end{eqnarray}

\section{Relative entropy in free bosonic theory}
\label{section:Examples}

In this section we are going to apply the formalism reviewed above to 
work out some new results for the (R\'enyi) relative entropy between eigenstates of the massless free bosonic field theory, 
whose Euclidean action is 
\begin{equation}
\mathcal{S}[\varphi] = \frac{1}{8 \pi} \int dz d\bar{z} \; \partial_z \varphi \partial_{\bar{z}} \varphi\,,
\end{equation}
which is a CFT with central charge $c=1$. 
In the following we will denote with $\phi$ and $\bar{\phi}$  the chiral and antichiaral component of the bosonic field, i.e. 
$\varphi(z,\bar z) = \phi (z) + \bar{\phi} (\bar{z})$.
We will only consider the case of $A$ being one interval of length $\ell$ embedded in a finite system of total length $L$
with periodic boundary conditions. 

\subsection{Relative entropy between the ground state and the vertex operator: $V_{\beta}$/GS}

The first case we study is the relative entropy between the ground state and the 
excited state generated by a vertex operator, which is a primary operator of the theory, defined as
\begin{equation}
V_{\alpha, \bar{\alpha}}\equiv :  e^{i (\alpha \phi + \bar{\alpha} \bar{\phi} )} : \, .
\end{equation}
We will focus on its chiral component (i.e. $\bar{\alpha}=0$), with conformal dimensions  $(h, \bar{h})= (\frac{\alpha^2}{2}, 0)$ and 
we will denote by $\rho_{V_{\alpha}}= {\rm Tr}_{\bar{A}} | V_{\alpha} \rangle \langle V_{\alpha} | $ the associated RDM.
This relative entropy has already been considered in Ref. \cite{lashkari2016}, but it is important to repeat the calculation here 
to set up the formalism and because we will need some informations from this calculation in  the following.

\begin{figure}
\centering
\centering
\subfigure
  {\includegraphics[width=0.45\textwidth]{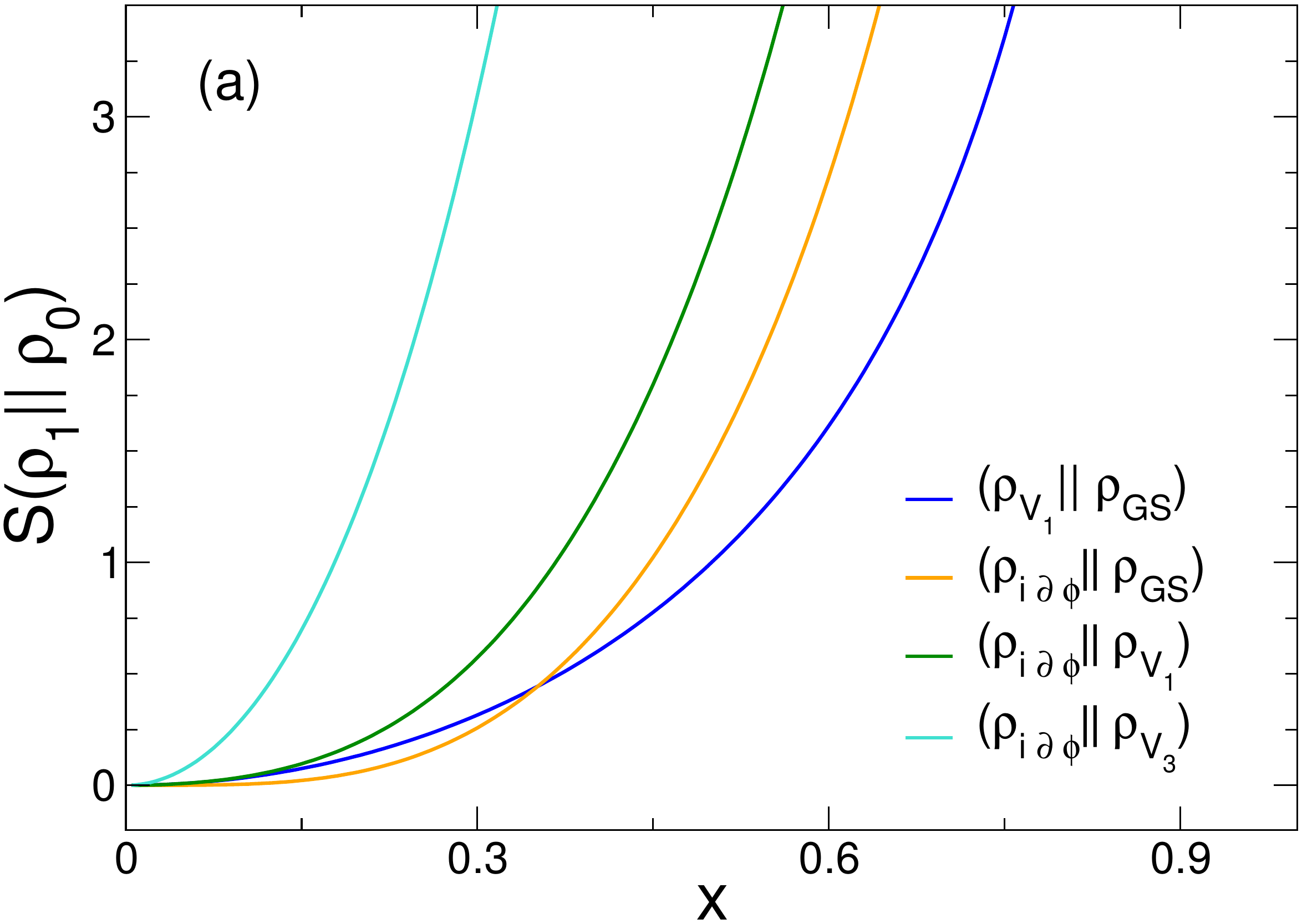}}
\subfigure
   {\includegraphics[width=0.45\textwidth]{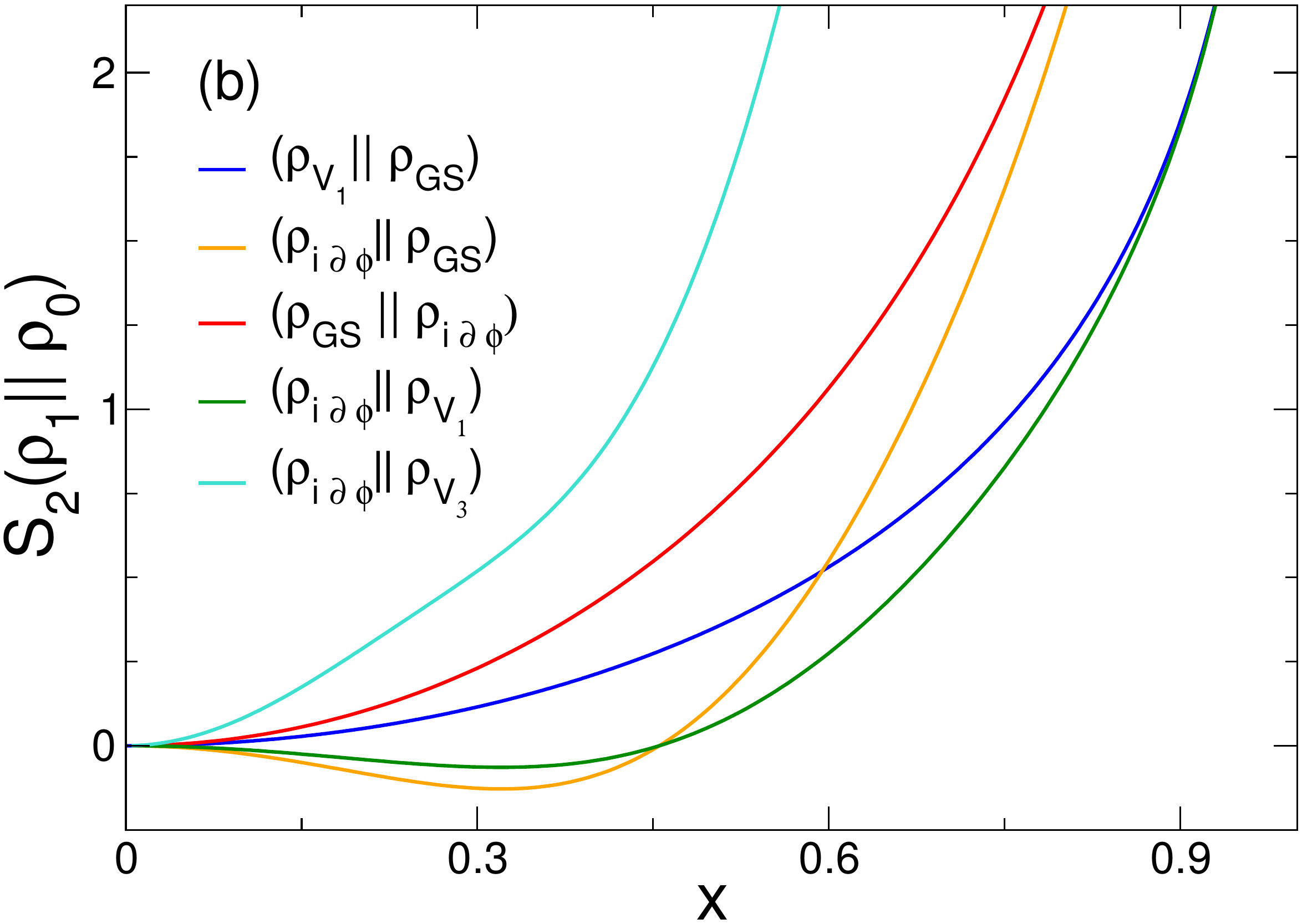}}
\subfigure
   {\includegraphics[width=0.45\textwidth]{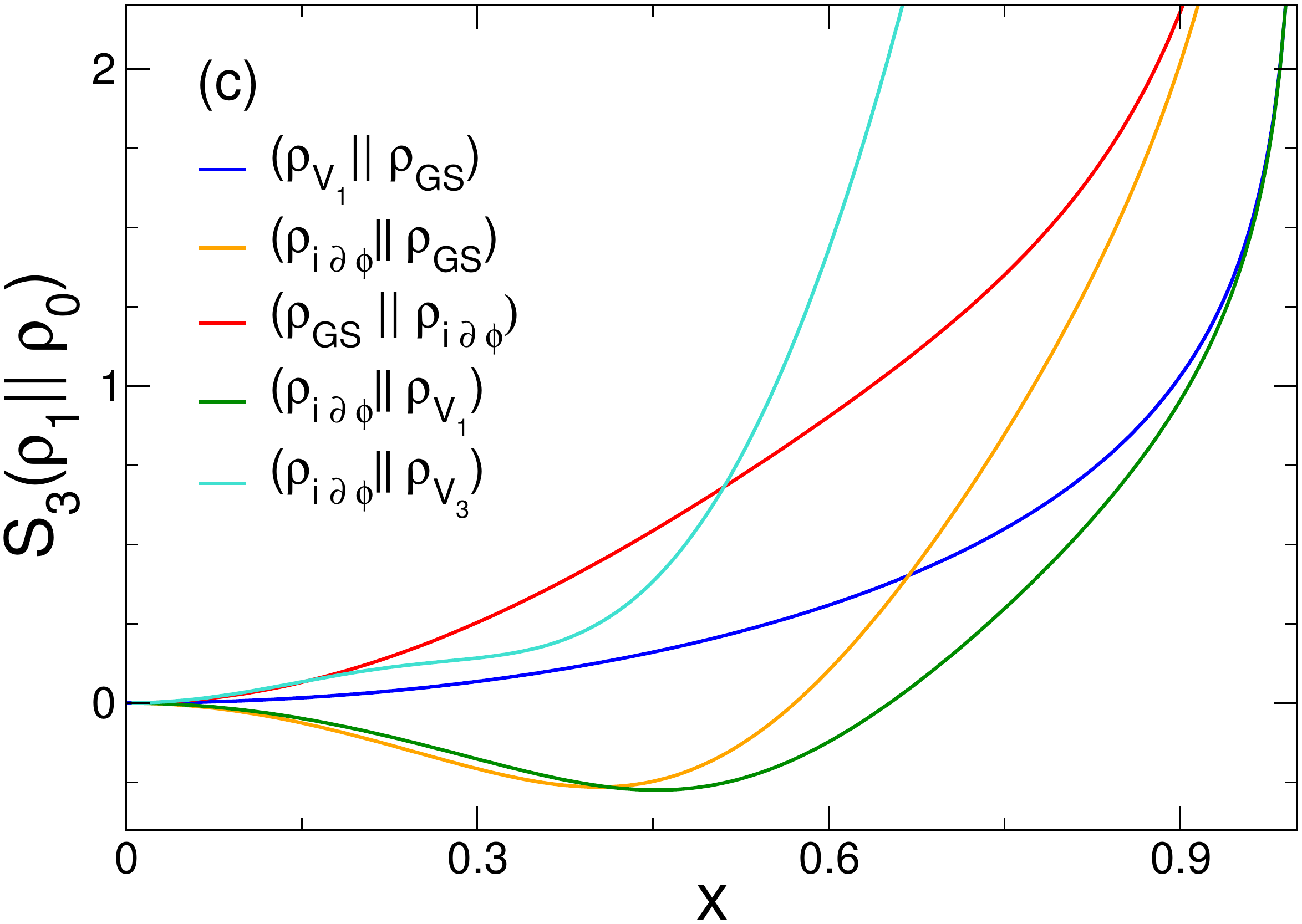}}
    \subfigure
  {\includegraphics[width=0.45\textwidth]{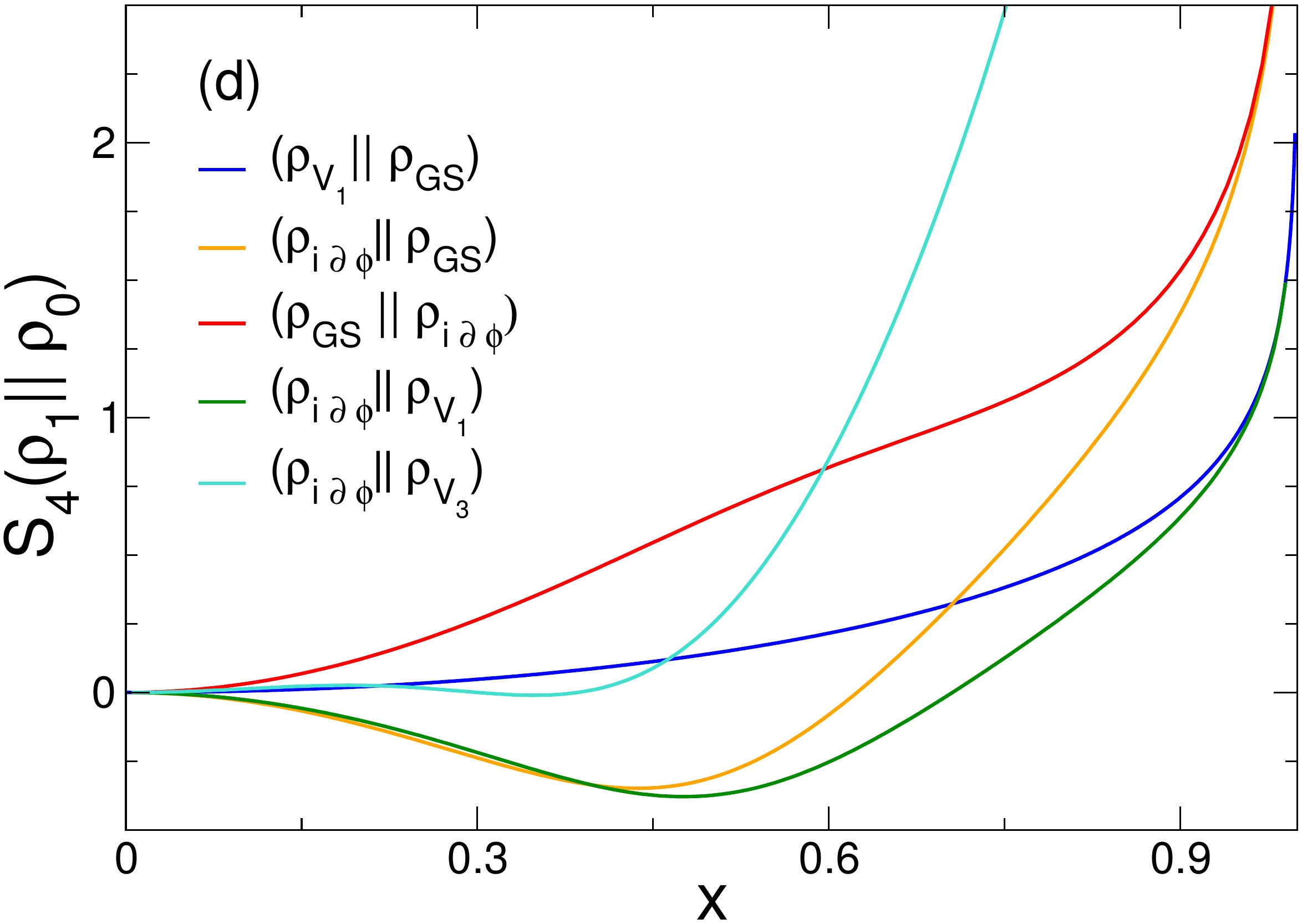}}
\caption{The CFT predictions for the R\'enyi relative entropies $S_{n} ( \rho_1 \| \rho_0 )$  as a function of $x= \ell/N$
for different values of $n =1, 2, 3, 4$ in panels (a), (b), (c) and (d) respectively. 
In each panel (at fixed $n$) we report the various states that we have considered in this paper, 
in order to compare the various results. 
Notice the non-positivity and non-monotonicity of some $S_{n} ( \rho_1 \| \rho_0 )$ for $n\neq1$.
 }
\label{relative_entropies}
\end{figure}

The $2n$-point correlation function of vertex operators on the complex plane is \cite{cft-book}
\begin{equation}
\langle \prod_{k} V_{\alpha_{k}} (w_k) \rangle = \prod_{k >i } (w_k - w_i)^{\alpha_k \alpha_i},
\end{equation}
and after the mapping to the variable $t$ (cf. \eqref{tdef}) it becomes ($t_{k i } \equiv t_{k,n}- t_{i,n}$)
\begin{equation}
\langle \prod_{k} V_{\alpha_{k}} (t_k) \rangle = \prod_{k >i } \left[ 2 \sin(t_{ki}/2) \right]^{\alpha_k \alpha_i}.
\label{Vat}
\end{equation}
Plugging this expression in (\ref{rhosigmaRel2}) and (\ref{sigmarhoRel2}), we derive the following results for the replicated 
relative entropies 
{\small
\begin{eqnarray*}
G^{(n)} ( \rho_{GS} \| \rho_{V_{\alpha}} ) &=&  
 \left[  n^{-(n-1)}   \left( \frac{\sin (\pi x)}{\sin (\pi x/n)} \right)^{n-1}  \prod_{m=1}^{n-2} \left( \frac{\sin^2 (\pi m/n)}{\sin(\pi (x+ m)/n) \sin (\pi (x-m)/n)  } \right)^{n-1-m}   \right]^{\alpha^2} ,
 \label{Va1}  \\
G^{(n)} ( \rho_{V_{\alpha}} \| \rho_{GS} ) &=& \left[ n^{n-1} \left( \frac{\sin (\pi x)}{ \sin (\pi x/ n)} \right)^{1-n}  \prod_{m=1}^{n-1} \left( \frac{\sin^2 (\pi m/n)}{\sin(\pi (x+ m)/n) \sin (\pi (x-m)/n)  } \right)^{m-n}   \right]^{\alpha^2},
\label{Va2}
\end{eqnarray*}
}
and making use of  identities
\begin{eqnarray}
\prod_{m=1}^{n-2} \left( \frac{\sin^2 (\pi m/n)}{\sin(\pi (x+ m)/n) \sin (\pi (x-m)/n)  } \right)^{n-1-m}   &=&  \left( \frac{ n \sin (\pi x/n)}{\sin (\pi x)} \right)^{n-2}, \\
\prod_{m=1}^{n-1} \left( \frac{\sin^2 (\pi m/n)}{\sin(\pi (x+ m)/n) \sin (\pi (x-m)/n)  } \right)^{n-m} &=&   \left( \frac{ n \sin (\pi x/n)}{\sin (\pi x)} \right)^{n},
\end{eqnarray}
they simplify to
\begin{equation}
G^{(n)} ( \rho_{GS} \| \rho_{V_{\alpha}} ) = G^{(n)} ( \rho_{V_{\alpha}} \| \rho_{GS} ) =  \left( \frac{\sin (\pi x)}{n \sin (\pi x/n)} \right) ^{\alpha^2}.
\end{equation}
Thus, it turns out that, for these specific operators, the  $G^{(n)}$ (and so the R\'enyi relative entropies $S_n$) are symmetric under  
exchange of the two reduced density matrices $\rho_{V_{\alpha}} \leftrightarrow \rho_{GS}$,
which, as already mentioned, is not true in general.
Of course, by replica limit $n\to1$, the same holds true also for the relative entropy  which is 
\begin{equation}
S (\rho_{V_{\alpha}} \| \rho_{GS} ) =  S ( \rho_{GS} \| \rho_{V_{\alpha}}  ) = \alpha^2 (1 - \pi x \cot (\pi x)).
\label{SV}
\end{equation}
In Figure \ref{relative_entropies} we plot the R\'enyi relative entropies $S_n(\rho_{V_{\alpha}} \| \rho_{GS} )$
as function of $x$ for $n=1,2,3,4$. They are all monotonous and positive function of $x$.

More generally, in \cite{lashkari2016} it has been shown that Eq. \eqref{SV} holds also for the relative entropy between two 
excited states of the form $V_{\alpha}|0\rangle$ with different charges $\alpha, \beta$, but with the 
replacement $\alpha\to\alpha-\beta$, i.e. \cite{lashkari2016}
\begin{equation}
S (\rho_{V_{\alpha}} \| \rho_{V_{\beta}} )= S (\rho_{V_{\beta}} \| \rho_{V_{\alpha}} ) =(\alpha-\beta)^2 (1 - \pi x \cot (\pi x)).
\end{equation}

\subsection{Relative entropy between the ground state and the derivative operator: $i \partial \phi$/GS}

Here we consider a more complicated case that has not yet been studied in the literature, namely the relative entropy 
of the excited state generated by $i\partial \phi$ (which is a primary operator of the theory with conformal 
dimensions $(h, \bar{h})= (1, 0)$) again with respect to the ground state.
We denote the RDM as $\rho_{i \partial \phi}= {\rm Tr}_{\bar{A}} | i \partial \phi \rangle \langle i \partial \phi | $.

The $2n$-point correlation function of $i \partial \phi$ in the complex plane is \cite{cft-book}
\begin{equation}
\langle \prod_{j=1}^{2n} i \partial \phi  ( z_j)\rangle_{\mathbb{C}} = \text{Hf} \left[ \frac{1}{z_{ij} ^2 } \right]_{ i, j \in [1,  2n]},
\end{equation}
where we denote with $ z_{ij} \equiv z_i - z_j$ and we introduced the Hafnian (Hf) as
\begin{equation}
\text{Hf}[A] \equiv \frac{1}{2^n n!}\sum_{p \in S_{2n}} \prod_{i =1}^n A_{p (2i - 1), p (2 i)}\,,
\end{equation}
with the sum being over all cyclic permutations.
This Hafnian can be expressed as a determinant using the following standard linear algebra identity
\begin{equation}
\text{Hf} \Big[ \frac{1}{z_{ij}^2} \Big] = \det \Big[ \frac{1}{z_{ij}} \Big].
\end{equation}

For the case of our interest, after mapping to the cylinder of length $2 \pi$ in the variable $t$ (cf. \eqref{tdef}), we have
\begin{equation}
\langle \prod_{j=1}^{2n} i \partial \phi  ( t_{j,n})\rangle_{\text{cyl}} 
= \frac{1}{4^n} \det \left[ \frac{1}{\sin(t_{ij}/2) } \right]_{i, j \in [1, 2n]}\,.
\end{equation}
Plugging this result into (\ref{rhosigmaRel2}) and (\ref{sigmarhoRel2}), we get
\begin{eqnarray}
G^{(n)} ( \rho_{GS} \| \rho_{i \partial \phi})  &=&  \Big(\frac{\sin \pi x}{n}\Big)^{2 (n-1)} \det\left[  \frac{1}{\sin(t_{ij}/2)} \right]_{i, j \in [1, 2(n-1)]}  ,\label{dphi1}  \\
G^{(n)} ( \rho_{i \partial \phi} \| \rho_{GS} ) &=&  \Big(\frac{\sin \pi x}n \Big)^{2 (1-n)} \left( \sin \frac{\pi x}{n} \right)^{-2} \left( \det \left[  \frac{1}{\sin (t_{ij}/2)} \right]_{i, j \in [1, 2n]} \right)^{-1}. \label{dphi2}
\end{eqnarray}

While the above functions are sufficient to determine the R\'enyi relative entropy of integer order, 
the relative entropies $S (\rho_{GS} \| \rho_{i \partial \phi})$ and $S (\rho_{i \partial \phi} \| \rho_{GS})$ are obtained from the 
analytic continuation of  (\ref{dphi1}) and (\ref{dphi2}) and taking the replica limit $n \to 1$. 
Such analytic continuations are however very difficult since the integer $n$ appear as the dimension of a matrix. 
Fortunately, for the determinant in \eqref{dphi2} the analytic continuation has been already worked out 
\cite{elc-13,CEL} and it is given by
\begin{equation}
\det \left[  \frac{1}{\sin (t_{ij}/2)} \right]_{i, j \in [1, 2n]}  = 4^n\frac{\Gamma^2 \left(  \frac{1+n + n \csc \pi x}{2}   \right)}{\Gamma^2 \left(  \frac{1 - n + n \csc \pi x}{2}  \right)  }.
\end{equation}
Thus the relative entropy can be straightforwardly computed obtaining 
\begin{equation}
S(\rho_{i \partial \phi} \| \rho_{GS})= 2 \left( \log (2  \sin (\pi x) )  + 1 - \pi x \cot( \pi x) + \psi_0 \left( \frac{\csc (\pi x)}{2}\right)   + \sin (\pi x)  \right),
\end{equation}
where $\psi_0 (z)$ is the digamma function. 
The expansion of this relative entropy for small $x$ agrees with the general result in \cite{ugajin2016}.

Finding instead the analytic continuation of (\ref{dphi1}) is much more complicated. 
The technical difficulty stems from the matrix in (\ref{dphi1}) having dimension  $n-1$ instead of $n$, 
an apparently innocuous change that alters completely the  structure of the eigenvalues as it could be verified by a 
direct inspection for small $n$. 
We mention that, in case one would be interested in an approximate estimate of this relative entropy, it 
is sufficient to employ a rational approximation for the analytic continuation as explained in Ref. \cite{dct-15}. 

In Figure \ref{relative_entropies} we plot the R\'enyi relative entropies $S_n(\rho_{GS} \| \rho_{i \partial \phi})$
and $S_n( \rho_{i \partial \phi}\|\rho_{GS})$ for $n=1,2,3,4$.
While $S_n(\rho_{GS} \| \rho_{i \partial \phi})$ for $n=2,3,4$ is always positive 
and monotonous, this is not the case for $S_n( \rho_{i \partial \phi}\|\rho_{GS})$ which takes negative values and it is 
non monotonous for $n\neq 1$ (for $n=1$ is always positive, as it should).
Although  $S_n(\rho_{GS} \| \rho_{i \partial \phi})$ is always positive and monotonous, its second derivative 
clearly changes sign as a difference compared to $S_n(\rho_{GS} \| \rho_{V_\alpha})$.

\subsection{Relative entropy between the vertex and the derivative operators: $i \partial \phi/V_{\beta}$}

We finally consider the relative entropy between two different excited states, associated to $i \partial \phi$ and $V_{\beta}$ respectively.
In this case the replicated function is given by Eq.  \eqref{2ndreplicaSrel-t}. 
This requires the calculation of the $2n$-point correlation function
\begin{equation}
\langle i \partial \phi (t_{1,n}) i \partial \phi (t'_{1, n}) \prod_{j = 2}^{n-1} V_{\beta} (t_{j, n}) V_{- \beta}(t'_{j, n}) \rangle_{\text{cyl}},
\end{equation}
entering in  $G^{(n)}(\rho_{i \partial \phi} \| \rho_{V_{\beta}} )$, cf. \eqref{2ndreplicaSrel-t}.

Noticing that
\begin{equation}
i \partial \phi (t) = \left( \frac{1}{\alpha} \frac{\partial}{\partial t} V_{\alpha} (t) \right) \bigg|_{\alpha=0},
\end{equation}
we can relate the desired correlation function to the derivative of the $2n$-point correlation function of vertex operators
in the following way
\begin{multline}
\langle i \partial \phi (t_{1,n}) i \partial \phi (t'_{1,n}) \prod_{j = 2}^{n} V_{\beta} (t_{j, n}) V_{- \beta}(t'_{j, n}) \rangle = \\  
- \frac{1}{\alpha^2}  \frac{\partial}{\partial t_{1, n}} \frac{\partial}{\partial t'_{1, n}} 
 \langle  V_{\alpha} (t_{1,n})  V_{- \alpha} (t'_{1, n}) \prod_{j = 2}^{n} V_{\beta} (t_{j, n}) V_{- \beta}(t'_{j,n}) \rangle \bigg|_{\alpha=0} .  
\end{multline}
At this point we only have to deal with the $2n$-point correlation function of vertex operators, which is given in \eqref{Vat}.
By simple algebra, we can rewrite 
\begin{equation}
\langle i \partial \phi (t_{1,n}) i \partial \phi (t'_{1,n}) \prod_{j = 2}^{n} V_{\beta} (t_{j, n}) V_{- \beta}(t'_{j, n}) \rangle = 
\frac{\tilde{C}_{\alpha,\beta} (n, x)}{4\sin^2 \left( \frac{\pi x}{n} \right) }
\langle \prod_{j = 1}^{n-1} V_{\beta} (t_{j, n}) V_{- \beta}(t'_{j, n}) \rangle,
\end{equation}
where we defined 
\begin{equation}
\tilde{C}_{\alpha,\beta} ( n, x) \equiv   - \frac{4\sin^2 \left( \frac{\pi x}{n} \right) }{\alpha^2} \partial_{t_{1, n}} \partial_{t'_{1, n}} C_{\alpha,\beta}  (n, x),
\end{equation}
and
\begin{multline}
C_{\alpha,\beta} (n, x) \equiv 
\langle V_{\alpha} (t_{1, n}) V_{-\alpha} (t'_{1, n}) \rangle \times  \\
\times \prod_{m=1}^{n-1} \langle V_{\beta} (t_{m, n}) V_{\alpha} (t_{1, n})\rangle \langle V_{-\beta} (t'_{m, n}) V_{-\alpha} (t'_{1, n}) \rangle \langle V_{\beta} (t_{m, n}) V_{-\alpha} (t'_{1, n}) \rangle \langle V_{- \beta} (t'_{m, n}) V_{\alpha} (t_{1,n})\rangle.
\end{multline}
The factor $4\sin^2 \frac{\pi x}{n}$ has been introduced for later convenience. 

In $\tilde{C}_{\alpha,\beta} (n, x)$ the derivatives give rise to  many different terms, but most of them vanish when considering 
the limit for $\alpha\to0$. The explicit calculation is long but straightforward and the final result is
\begin{eqnarray}
\tilde{C}_{\alpha=0,\beta}(n, x)=   1 - \beta^2\sin^2 \left(\frac{\pi x}{n} \right) \left( \sum_{k=1}^{n-1} \cot \frac{\pi}{n}(x + k) \right)  \left( \sum_{l=1}^{n-1} \cot \frac{\pi}{n} (-x + l) \right) .
\end{eqnarray}

We now have all the needed correlations for the R\'enyi relative entropy $S_n( \rho_{i \partial \phi} \| \rho_{V_{\beta}})$ 
(or its exponential $G^{(n)} ( \rho_{i \partial \phi} \| \rho_{V_{\beta}} )$).
Plugging these correlations into \eqref{2ndreplicaSrel-t}, we have
\begin{multline}
G^{(n)} ( \rho_{i \partial \phi} \| \rho_{V_{\beta}} ) = n^{(\beta^2 - 2)(1-n)} \frac{\tilde{C}_{\alpha=0,\beta}(n, x)}{ 4\sin^2 \left( \frac{\pi x}{n} \right) }
\\ \times
\frac{\langle \prod_{k=1}^{n-1} V_{\beta} (t_{k,n}) V_{-\beta}(t'_{k,n}) \rangle_{\text{cyl}} \langle i \partial \phi (t_{1,1}) i \partial \phi (t'_{1,1}) \rangle_{\text{cyl}}^{n-1}  }{
\langle \prod_{k=0}^{n-1} i \partial \phi (t_{k,n})  i \partial \phi (t_{k, n}) \rangle_{\text{cyl}} \langle V_{\beta}(t_{1,1}) V_{-\beta}(t_{1,1}) \rangle_{\text{cyl}}^{n-1} }.
\end{multline}
Finally, using the explicit expressions for all the correlation functions (which are known from previous cases), we  get
\begin{multline} \label{Gnac_phiV}
G^{(n)} ( \rho_{i \partial \phi} \| \rho_{V_{\beta}} ) =  \\
\tilde{C}_{\alpha=0,\beta}(n, x)
\left( \frac{\sin (\pi x) }{n\sin \left( \frac{\pi x}{n} \right) } \right) ^{ \beta^2}
\left(\frac{\sin (\pi x)}n  \right)^{2 (1-n) } \frac1{ 4^n\sin^2\left( \frac{\pi x}{n} \right)}
\frac{\Gamma^2 \left(  \frac{1 - n + n \csc \pi x}{2}   \right)}{ \Gamma^2 \left(  \frac{1 + n + n \csc \pi x}{2}  \right)  } .
\end{multline}
This can be rewritten in the suggestive form 
\begin{equation}
G^{(n)} ( \rho_{i \partial \phi} \| \rho_{V_{\beta}} ) =  \tilde{C}_{\alpha=0,\beta}(n, x)
G^{(n)} ( \rho_{i \partial \phi} \| \rho_{GS} ) G^{(n)} ( \rho_{GS} \| \rho_{V_{\beta}} ) ,
\end{equation}
which shows that $G^{(n)} ( \rho_{i \partial \phi} \| \rho_{V_{\beta}} )$ is the product of two  $G^{(n)}$ 
of  $\rho_{i \partial \phi}$ or $\rho_{V_{\beta}}$ with respect to the ground state 
times an ``interaction term'' given by $\tilde{C}_{\alpha=0,\beta}(n, x)$.

Now in order to take the derivative with respect to $n$ of \eqref{Gnac_phiV} and take the replica limit for the relative entropy, 
we would need the analytic continuation to $n \in \mathbb{C}$ of the following finite sum
\begin{equation} \label{sumcot}
\sum_{k=1}^{n-1} \cot \frac{\pi}{n} (z + k).
\end{equation} 
This is easily done by using an integral representation of the cotangent an inverting the sum with the integral. 
However, this is not necessary because in the replica limit \eqref{rep_Srel}, these contributions are multiplied by a term 
vanishing for $n \to 1$.
Therefore it is straightforward to derive an analytic expression for the relative entropy, which ultimately reads
\begin{multline} \label{Srel_phiGS}
S  (\rho_{i \partial \phi} \|  \rho_{V_{\beta}} )= (2 + \beta^2)[1 - \pi x  \cot (\pi x) ]+ 2 \log ( 2\sin(\pi x) ) + 2 \psi_0 \left( \frac{\csc (\pi x)}{2}  \right) + 2 \sin (\pi x)= \\=
S  (\rho_{i \partial \phi} \|  \rho_{GS} )+S  (\rho_{GS} \|  \rho_{V_{\beta}} ),
\end{multline}
i.e. it is just the sum of the relative entropies of the two operators with respect to the ground state given that the ``interaction term'' 
$\tilde{C}_{\alpha=0,\beta}(n, x)$ vanishes in the replica limit. 

The R\'enyi relative entropies $S_n(\rho_{i \partial \phi} \|  \rho_{V_{\beta}} )$ for $n=1,2,3,4$ are reported for $\beta=1,3$ 
in the four panels of Figure \ref{relative_entropies}.
As it should, the relative entropy $S_1$ is always positive and also monotonous. 
For $n\neq1$ we have instead a more complicated behavior. 
Indeed $S_n(\rho_{i \partial \phi} \|  \rho_{V_{\beta}} )$ can be either positive or negative and the range of negativity 
depends on the values of both $n$ and $\beta$.
It is easy to see numerically that for any integer $n$, it exists a critical value $\beta_c(n)$ such that for $\beta>\beta_c(n)$, 
$S_n(\rho_{i \partial \phi} \|  \rho_{V_{\beta}} )$ is always positive, but not always monotonous. 

We mention that there are no conceptual difficulties for the calculation of 
$G^{(n)} ( \rho_{V_{\beta}} \| \rho_{i \partial \phi} )$ for finite integer $n$.
However, the computation requires to take $2(n-1)$ derivatives and therefore it is rather involved, 
especially  if one desires a closed form valid for arbitrary $n$.


 \begin{figure}[t]
 \centering
 \subfigure
   {\includegraphics[width=0.45\textwidth]{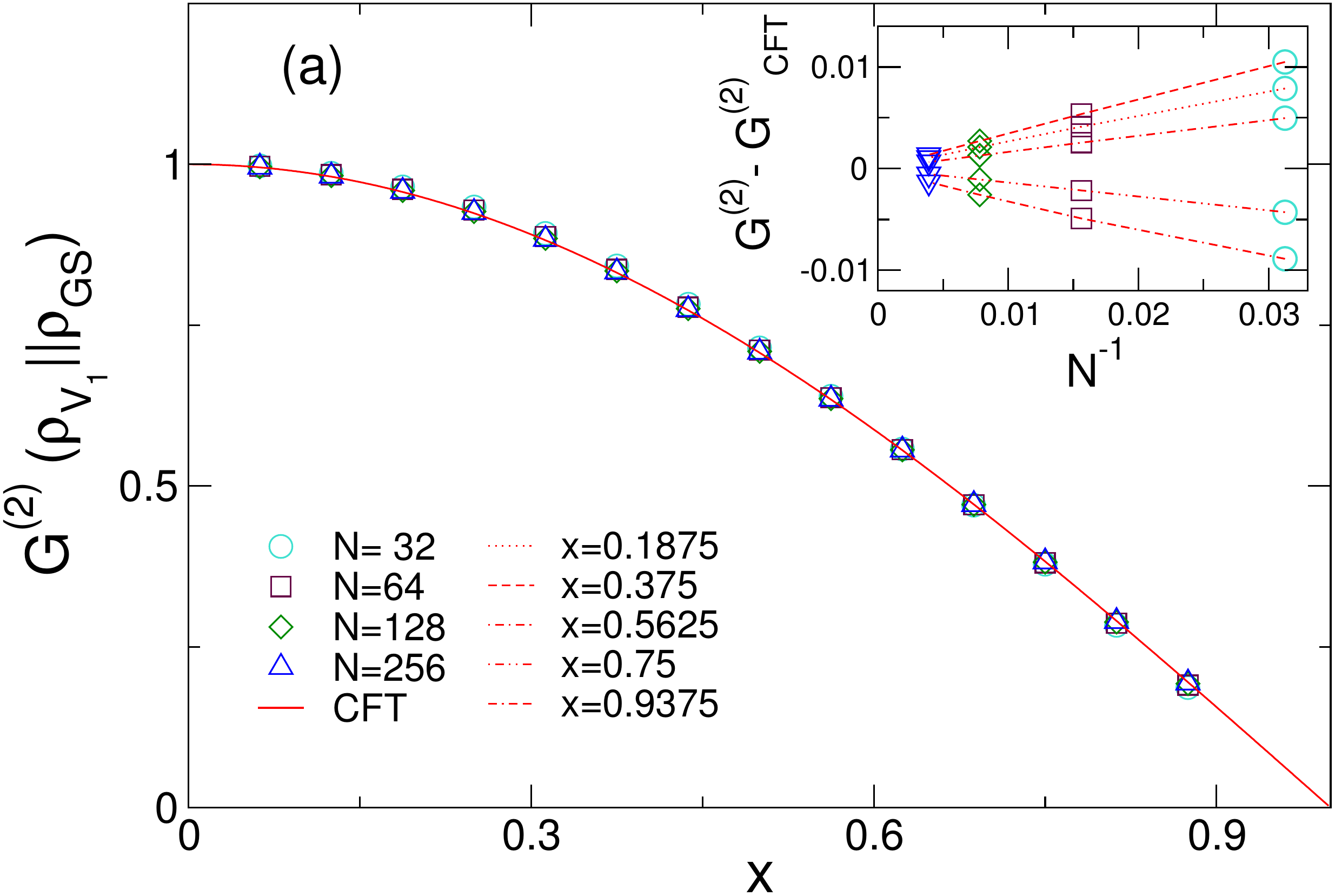}}
 \subfigure
   {\includegraphics[width=0.45\textwidth]{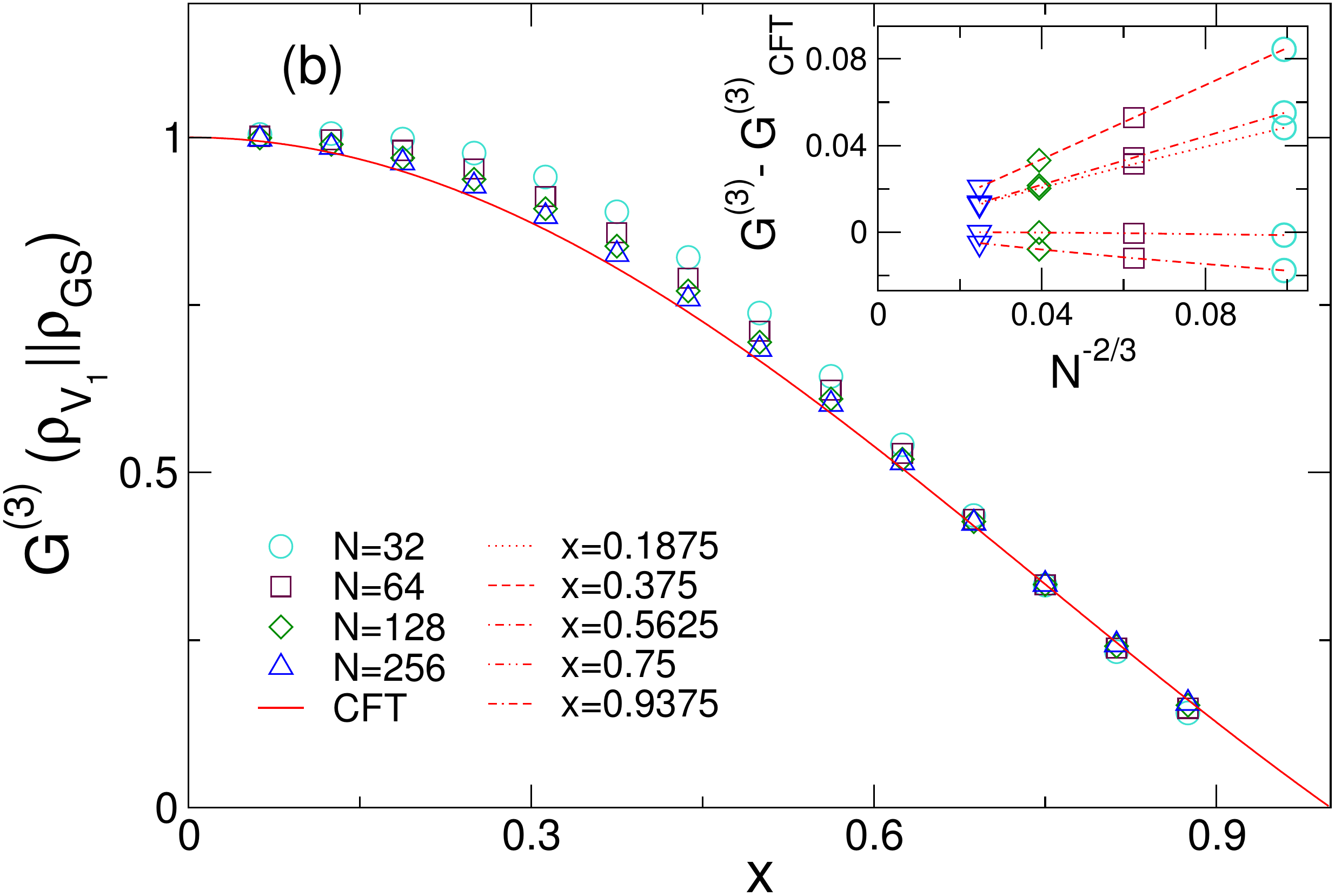}}
 \subfigure   
    {\includegraphics[width=0.45\textwidth]{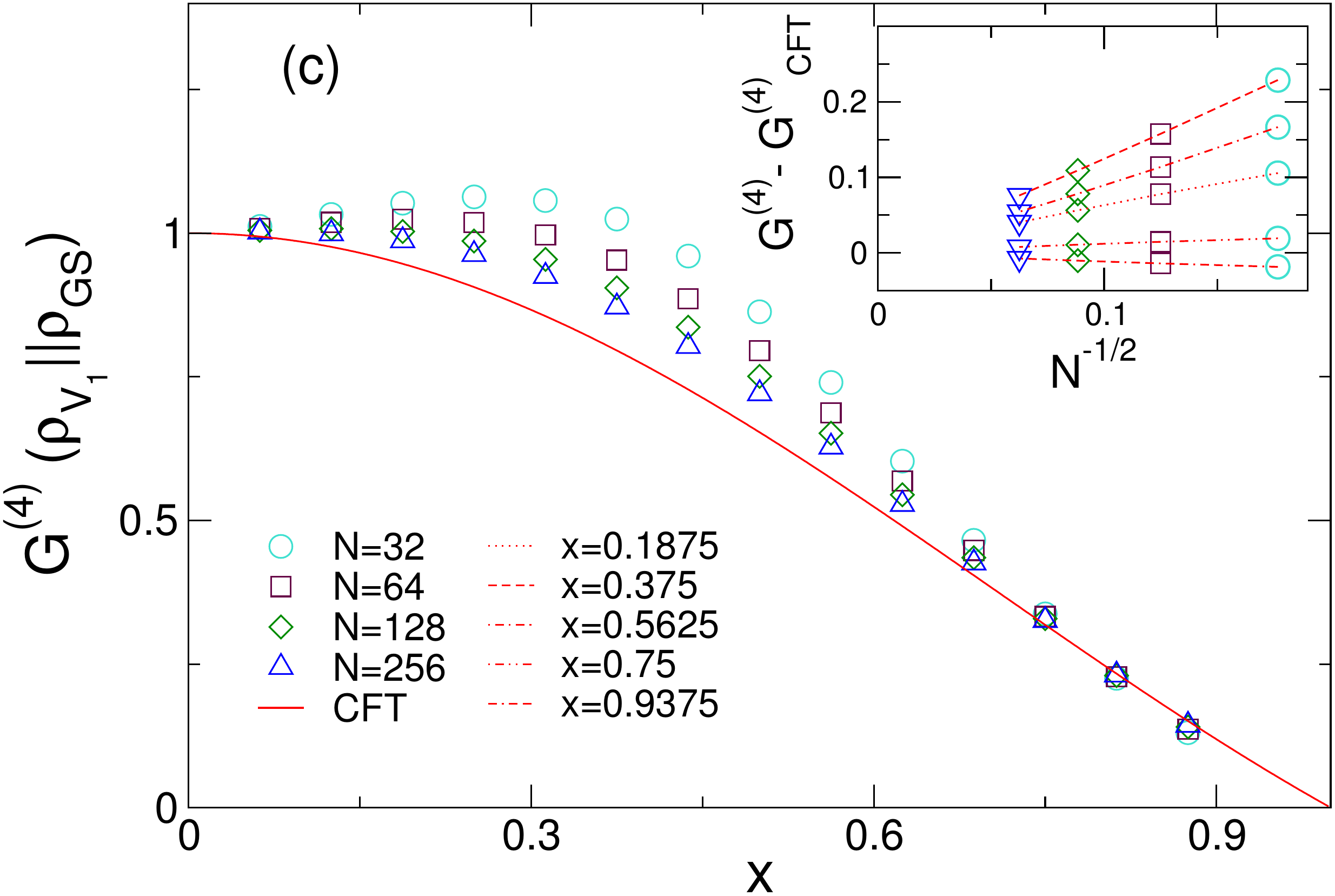}}
 \subfigure   
    {\includegraphics[width=0.45\textwidth]{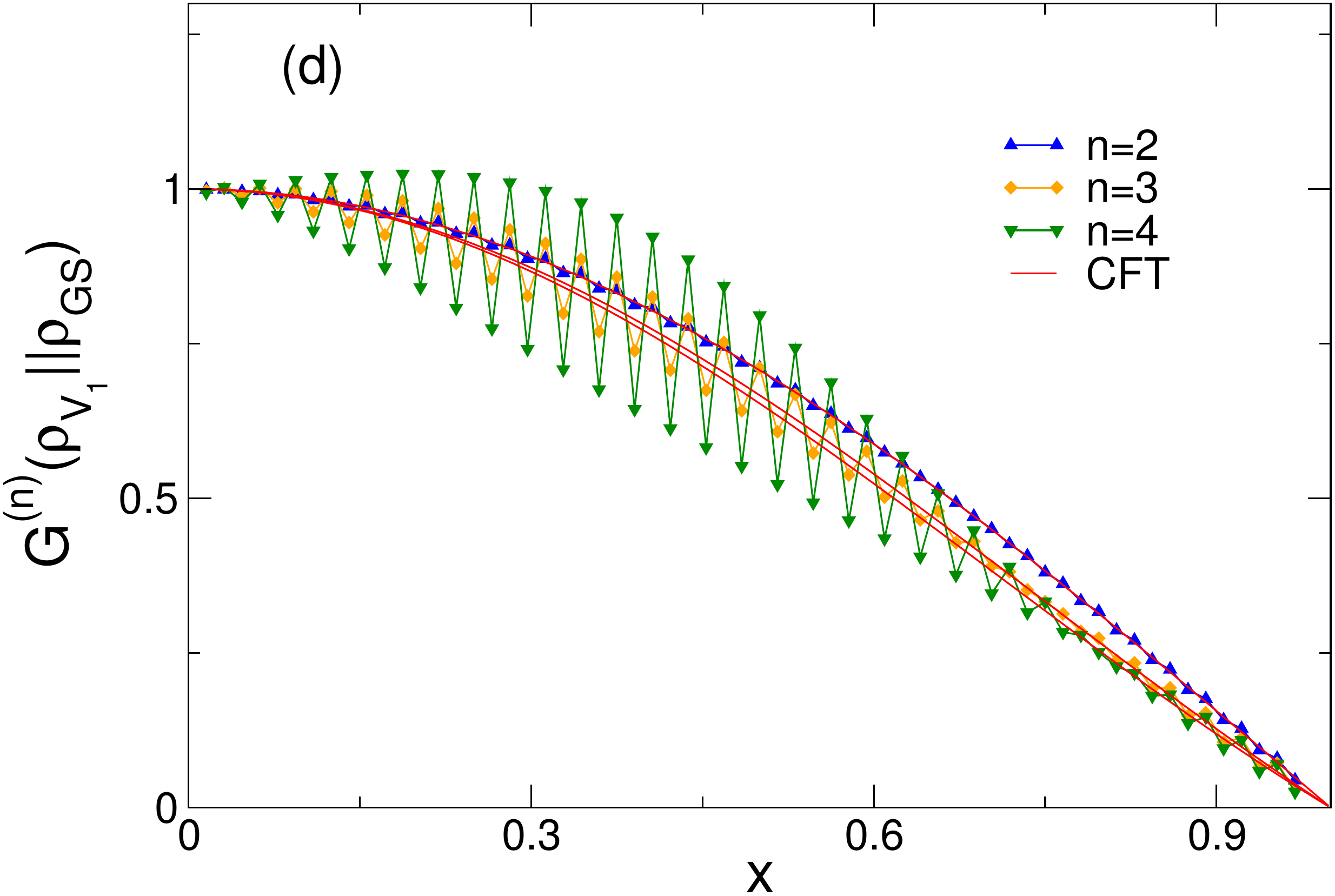}}
 \caption{The quantity $G^{(n)} ( \rho_{V_{1}} \| \rho_{GS} )$ as a function of $x= \ell/N$ for different values of $n \; (=2, 3, 4$ in panels (a), (b) and (c) respectively) plotted only for even values of $\ell$. 
Different symbols correspond to different system sizes and the red curve is the CFT prediction. 
The insets show the difference between the data in the XX model and the CFT prediction, against the leading scaling corrections  
$N^{-2/n}$. Each line corresponds to a given value of $x$. 
Panel (d): highlight on the oscillations found with the parity of the block's length for a single chain of total length $N=64$. 
Symbols of different colors correspond to different values of $n$ and the red curves are the CFT predictions. 
}
 \label{Gn_VGS}
 \end{figure}

\section{The XX spin-chain as a test of the CFT predictions}
\label{section:comparison}

\subsection{The model and its spectrum} 

The goal of this section is to check the validity of the formulas presented in the previous section in a lattice model, 
a fundamental test that has not yet been performed in the literature. 
We consider the easiest model to study the entanglement properties, namely  the XX spin-chain defined by the hamiltonian
\begin{equation} \label{xxmodel}
H_{XX}= - \frac{1}{4} \sum_{m= 1}^N \left[  \sigma^x_m \sigma^x_{m+1} + \sigma^y_m \sigma^y_{m+1} - h\sigma^z_m\right] ,
\end{equation}
where $\sigma^{x,y,z}_m$  are the Pauli matrices acting on the $m$-th spin and we assume periodic boundary 
condition.
By a Jordan-Wigner transformation, the spin hamiltonian is mapped into a free fermionic one of the form 
\begin{equation}
H_{XX}= - \frac{1}{2} \sum_{m=1}^N \left[  c_m^{\dagger} c_{m+1} + c_{m+1}^{\dagger} c_m +2 h \Big(c_m^{\dagger} c_m-\frac12\Big)\right],
\label{Hxxf}
\end{equation}
where $c^\dagger_m$ and $c_m$ are creation and annihilation operators at the site $m$.
The ground-state is a partially filled Fermi sea with Fermi-momentum $k_F =\arccos |h|$ and the single-particle dispersion relation 
$\epsilon_k=|\cos k-h|$, which can be linearized close to the two Fermi points $k=\pm k_F$, ending up with the two 
chiral components of a massless Dirac fermion which describes the low energy physics of the model.
Via bosonization this is nothing but the massless boson considered in the previous section in CFT formalism.


 \begin{figure}
 \centering
 \subfigure
   {\includegraphics[width=0.45\textwidth]{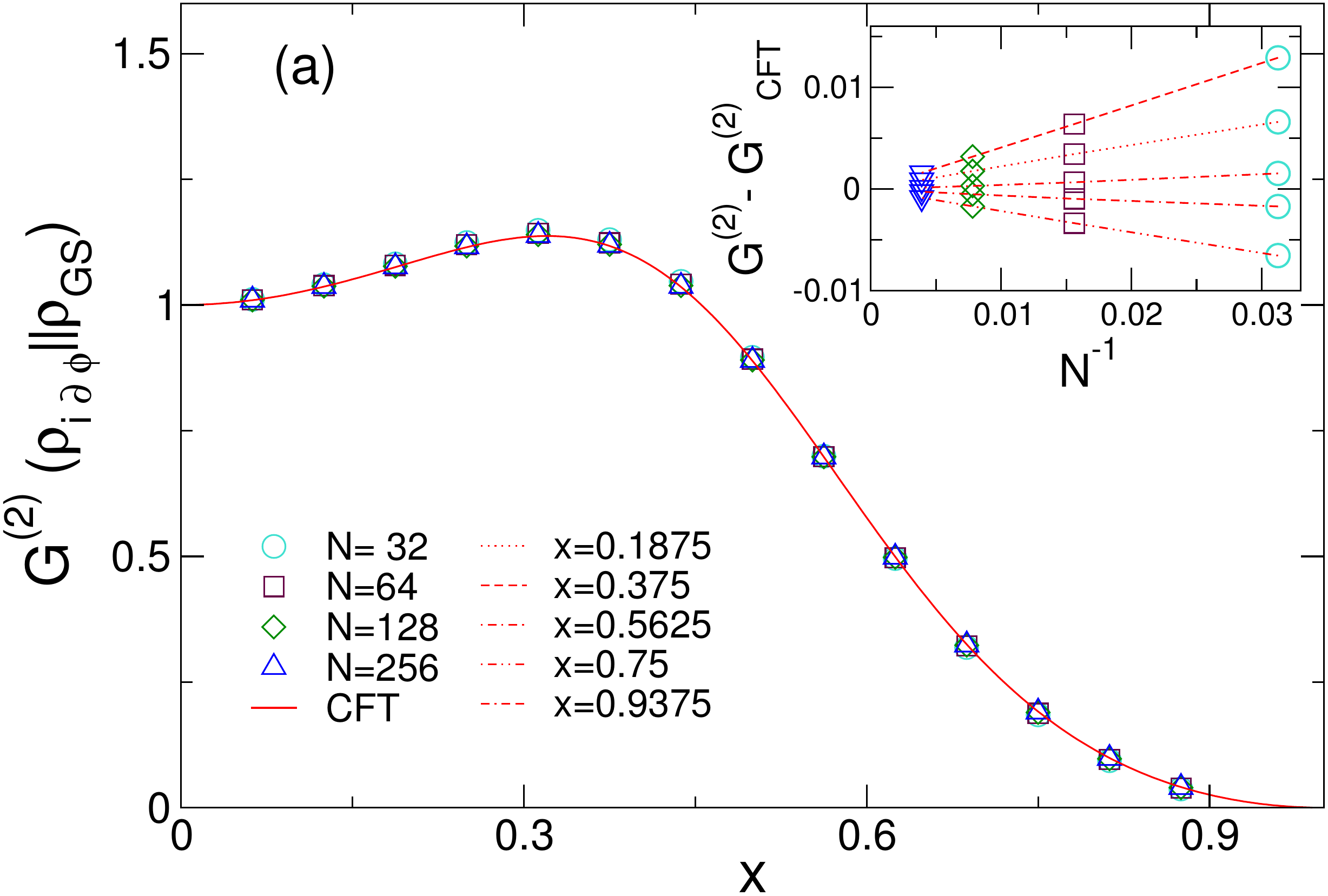}}
 \subfigure
   {\includegraphics[width=0.45\textwidth]{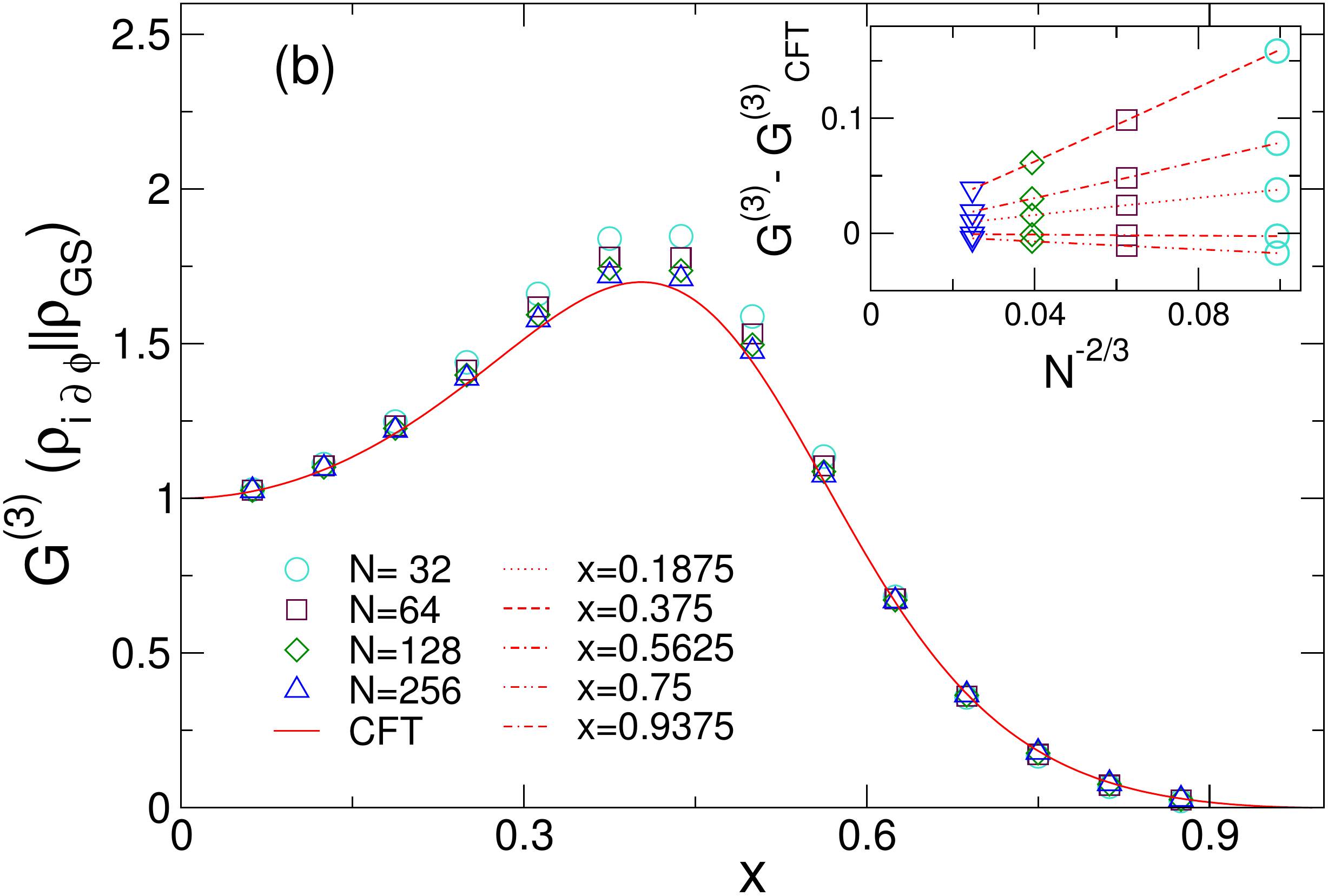}}
 \subfigure   
    {\includegraphics[width=0.45\textwidth]{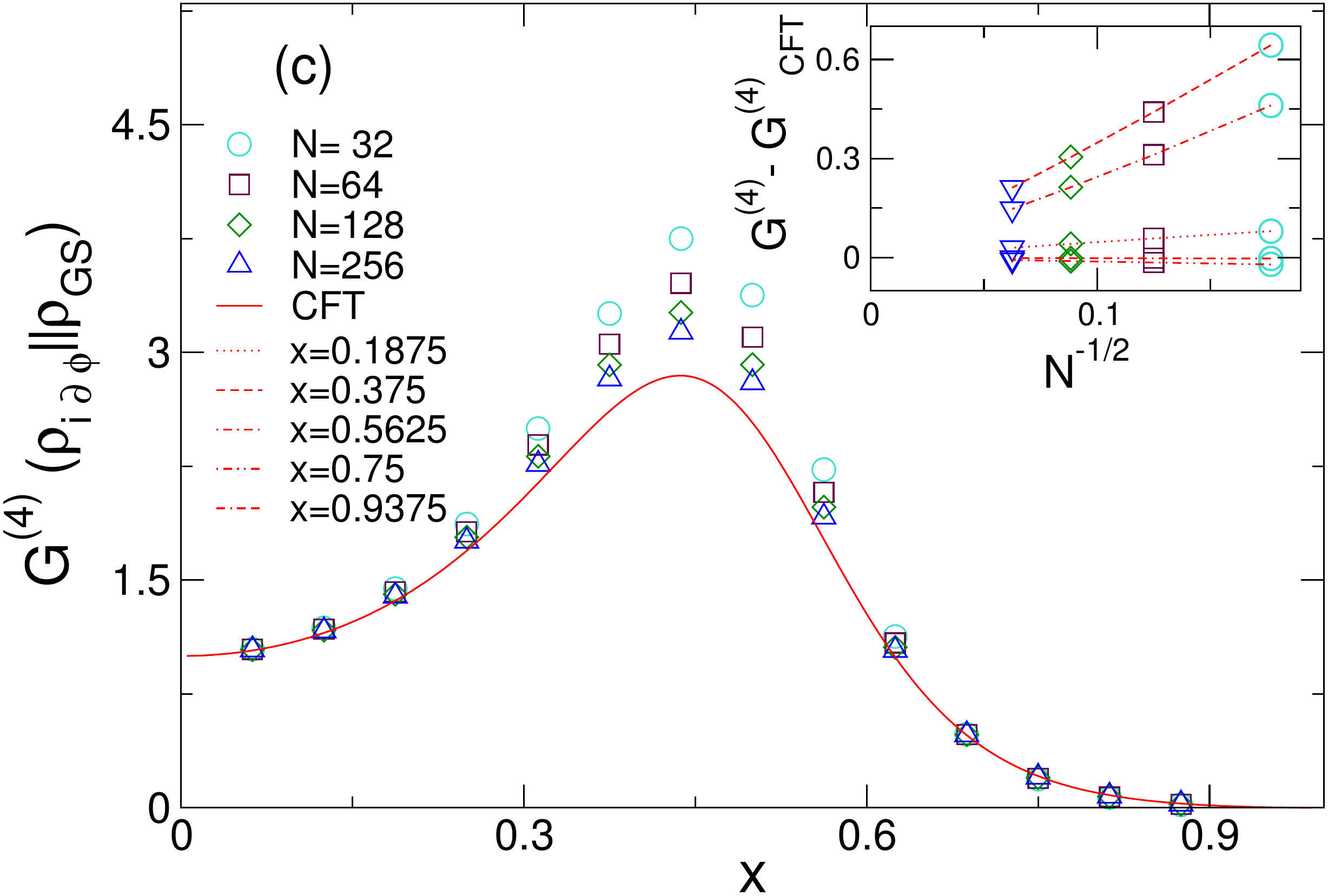}}
  \subfigure
   {\includegraphics[width=0.45\textwidth]{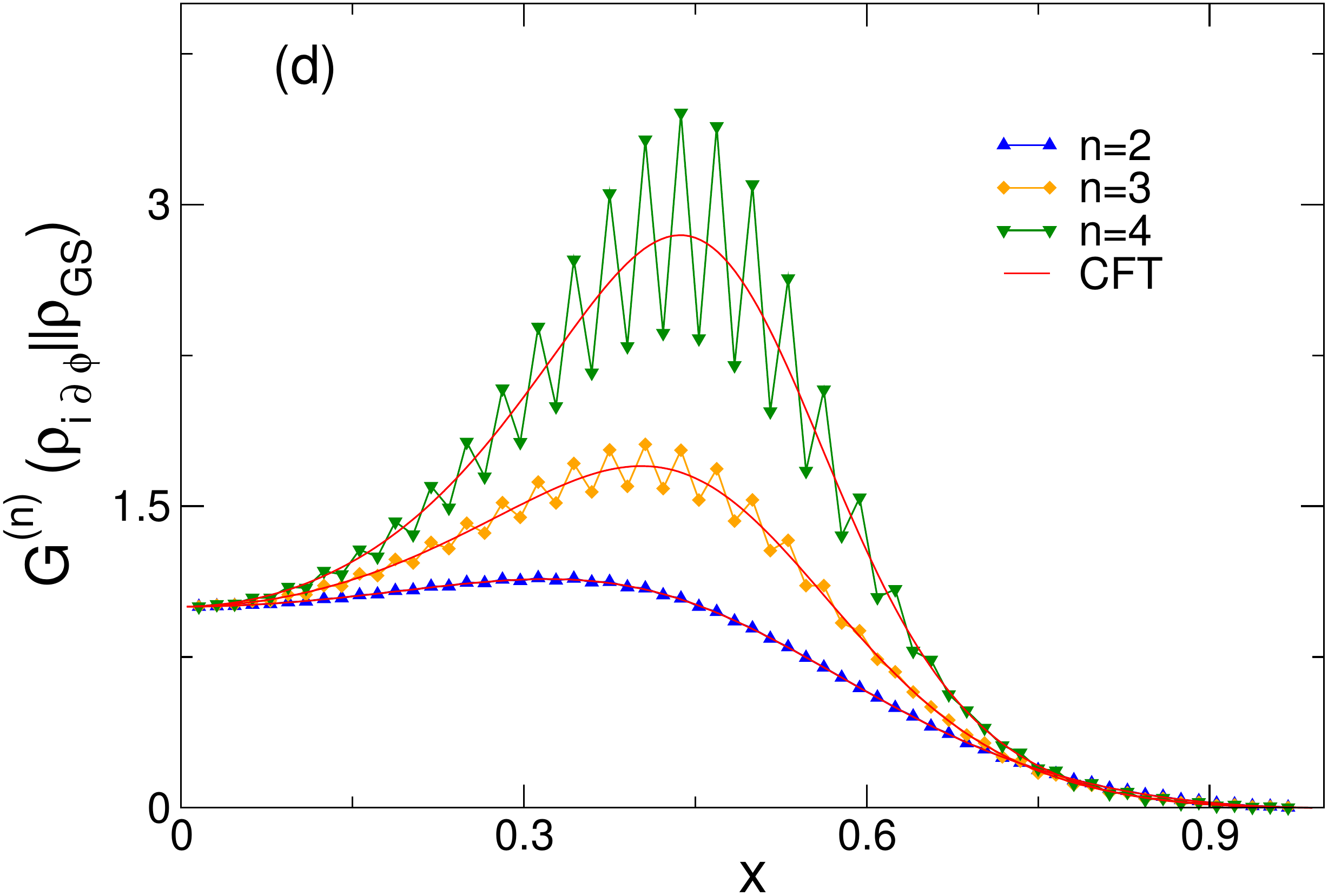}}   
 \caption{The quantity $G^{(n)} ( \rho_{i \partial \phi} \| \rho_{GS} )$ as a function of $ x= \ell/N$ for different values of $n$. 
 The description of the Figure is the same as in Fig. \ref{Gn_VGS}.}
 \label{Gn_phiGS}
 \end{figure}

Each eigenstate of the hamiltonian is in correspondence with a set of momenta $K$, 
corresponding to the occupied states
\begin{equation}
\prod_{k \in K} c^{\dagger}_k |0\rangle,
\end{equation}
with, e.g., the ground-state corresponding to $K$ being the set of all momenta with absolute value smaller than $k_F$.
Low-lying excited states are obtained by removing/adding some particles in momentum space close to the Fermi sea
and they can be written as a sequence of  creation/annihilation operators applied to the ground state.
These low lying excited states in the continuum limit can be put in one to one correspondence with the action of CFT primary operators 
onto the vacuum. 
A very detailed discussion on this correspondence between lattice and CFT excitations can be found, e.g., in Ref. \cite{sierra2012}, 
we just mention here the two states of our interest. 
Eigenstates in the middle of the spectrum have been studied in \cite{afc-09}.

We will only consider a vanishing external field $h=0$ which corresponds to a half-filled Fermi sea with $k_F=\pi/2$.
Furthermore, for simplicity, we focus on chains of length $N$ multiples of $4$, that at half-filling has $n_F=N/2$ fermions. 

The CFT state generated by a vertex operator $V_{\beta=1}|0\rangle$ corresponds in the XX chain to a hole-type excitation,
 i.e. the state \cite{sierra2012}
\begin{equation}
b_{ \frac{n_F}{2} -1 } |n_F \rangle,
\label{s1}
\end{equation}
where $|n_F \rangle$ is the ground state of the half-filled in fermionic model.
The primary operator $\left( i\partial \phi \right) $ is instead associated to the particle-hole excitation \cite{sierra2012}
\begin{equation}
b_{ \frac{n_F}{2} -1 } b^{\dagger}_{ \frac{n_F}{2} +1 } | n_F \rangle.
\label{s2}
\end{equation}


 \begin{figure}[t]
 \centering
 \subfigure
   {\includegraphics[width=0.45\textwidth]{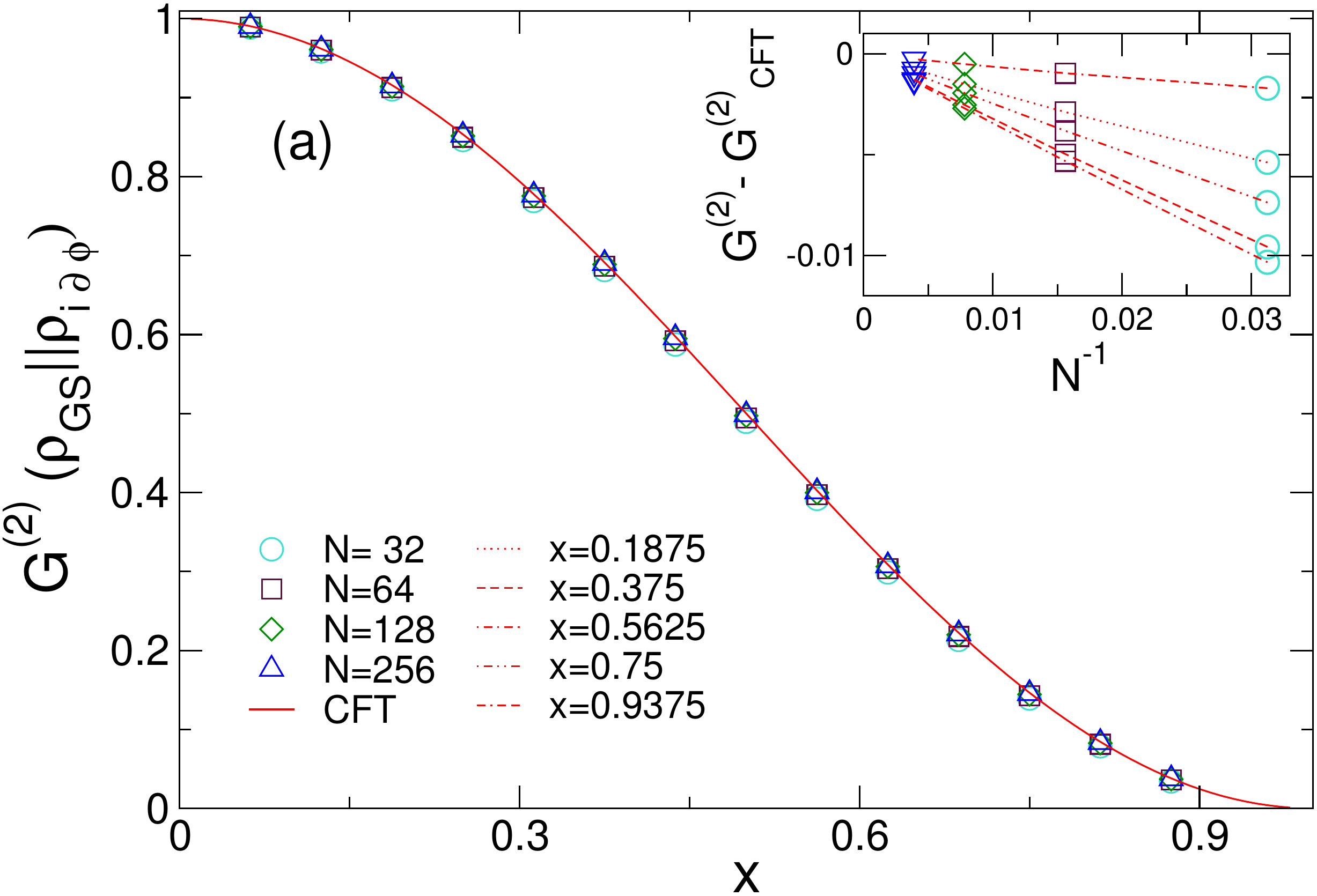}}
 \subfigure
   {\includegraphics[width=0.45\textwidth]{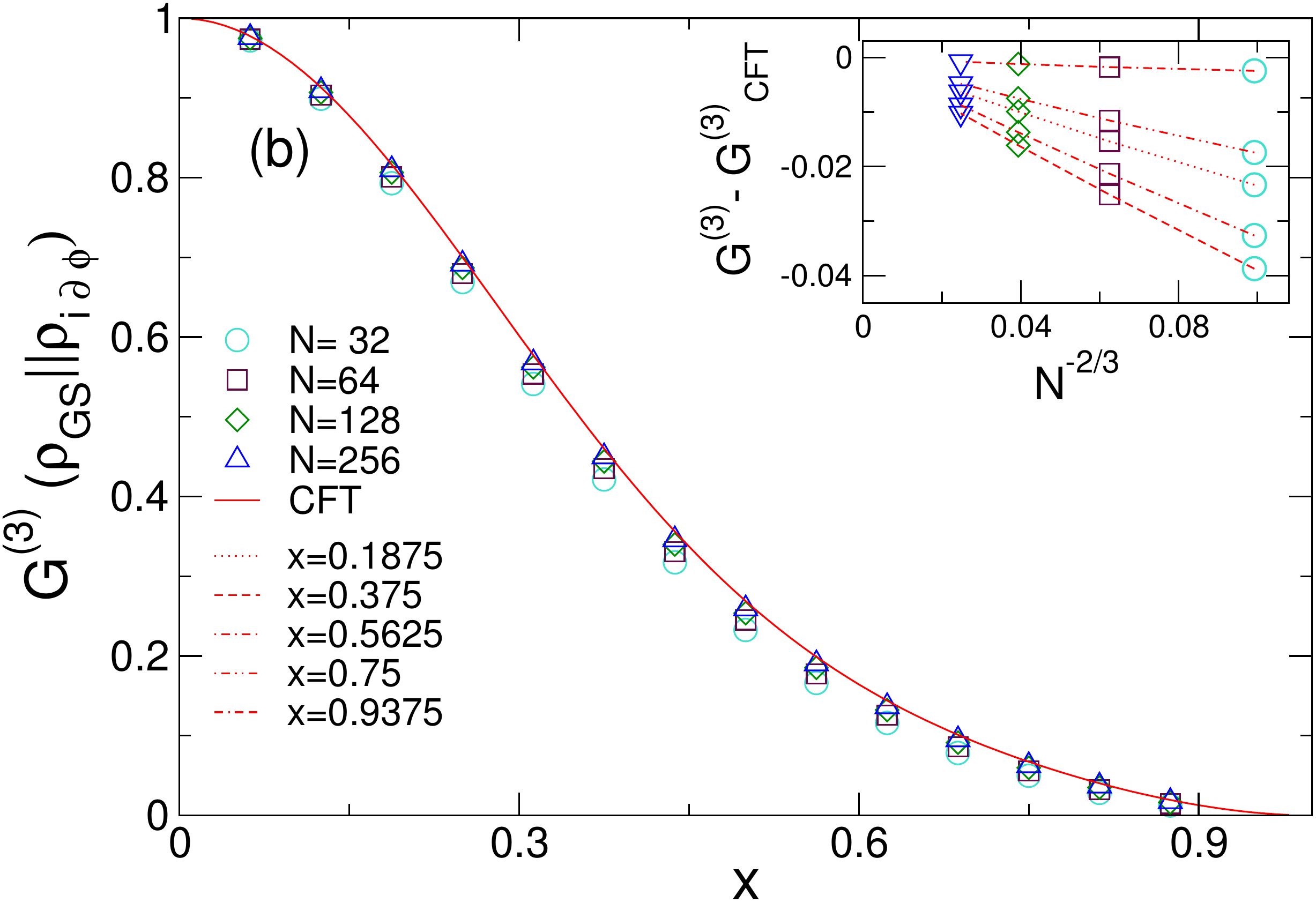}}
 \subfigure   
    {\includegraphics[width=0.45\textwidth]{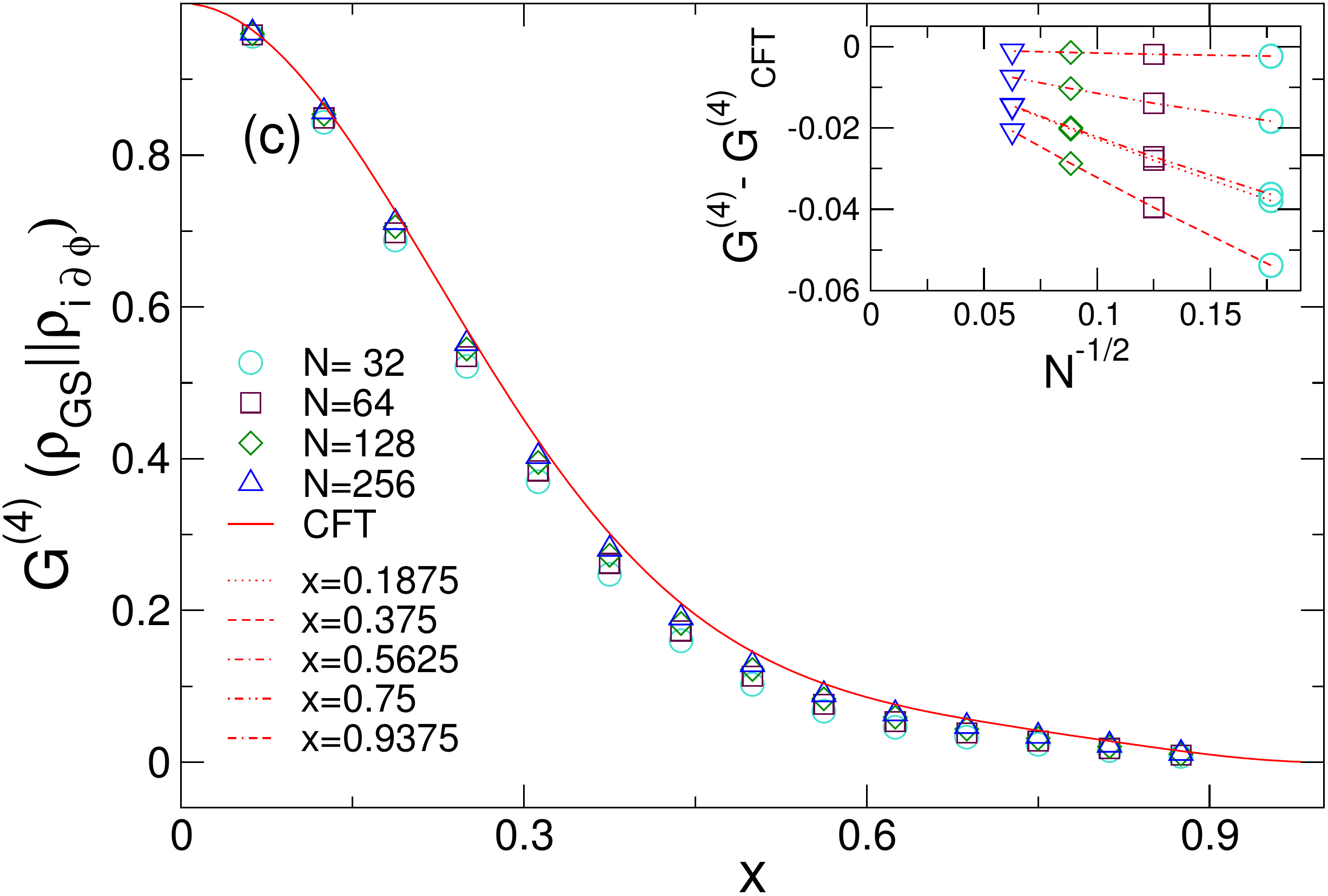}}
 \subfigure   
    {\includegraphics[width=0.45\textwidth]{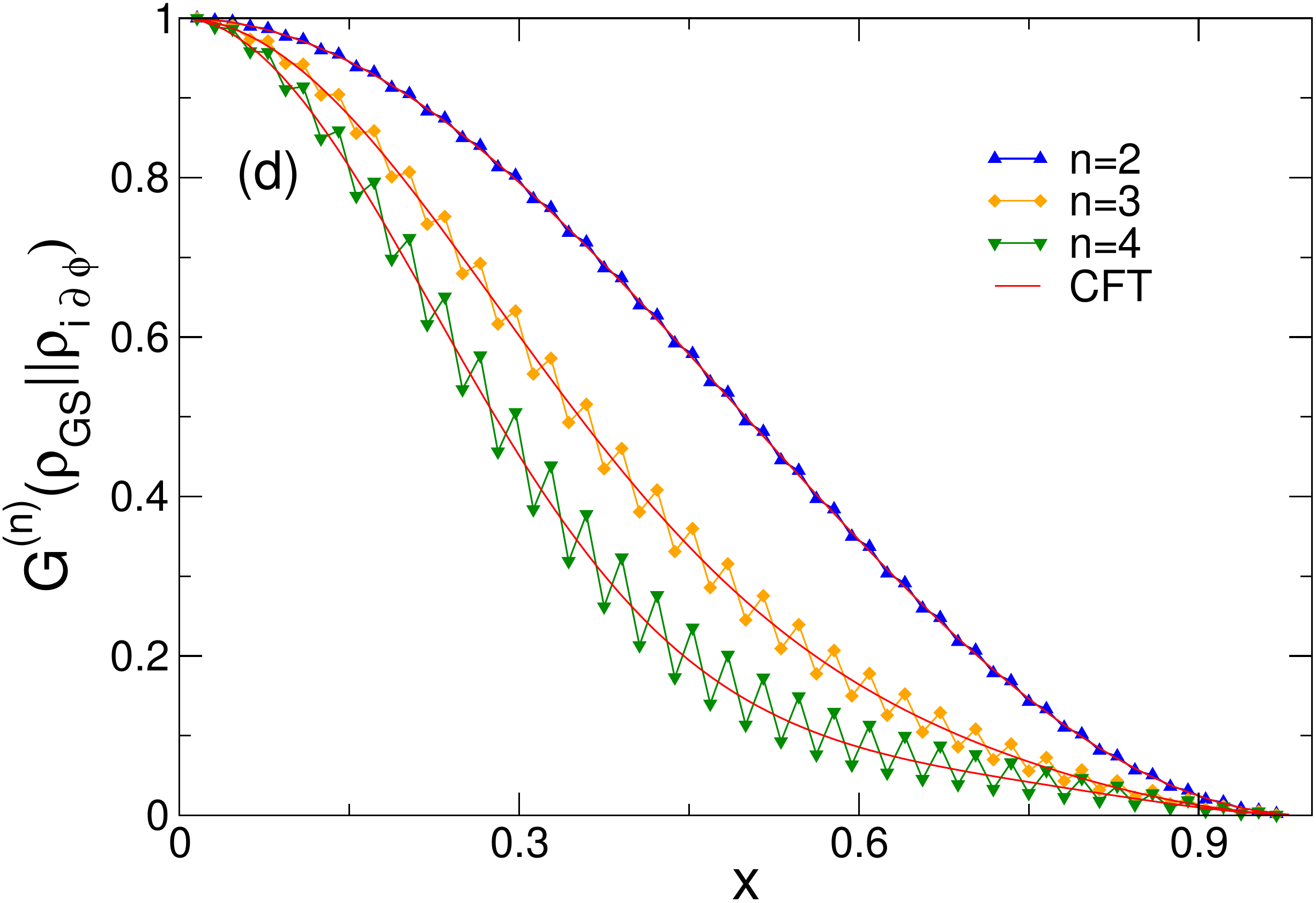}}
 \caption{The quantity $G^{(n)} ( \rho_{GS} \| \rho_{i \partial \phi} )$ as a function of $ x= \ell/N$ for different values of $n$. 
  The description of the Figure is the same as in Fig. \ref{Gn_VGS}.}
 \label{Gn_GSphi}
 \end{figure}

\subsection{Relative entropies and replicas}

Let us denote by $|\{k\}\rangle$ a generic eigenstate of the XX spin chain in which $\{k\}$ stands for the set of occupied 
single-particle levels. By Wick theorem, it is easy to show that the reduced density matrix of a block of $\ell$
contiguous sites can be written as \cite{vlrk-03,peschel2001,peschel2003,pe-09}
\begin{equation}
\rho_{\cal A}=\mathcal{K} \; e^{- H_A},
\label{gau}
\end{equation}
where $\mathcal{K}$ is a normalization constant and $H_A$ the modular (or entanglement) Hamiltonian that 
for Gaussian states takes the form
\begin{equation}
H_A = \sum_{ij} h_{ij} c_i^{\dagger} c_j\,.
\end{equation}
This modular Hamiltonian is related to the correlation matrix restricted to the block $A$ (with elements 
$[C_A]_{nm}= \langle \{k\} |c_m^\dagger c_n| \{k\} \rangle$ with $n,m\in A$)
as \cite{peschel2003}
\begin{equation}
h= \ln(C_A^{-1}-1).
\end{equation}
Denoting by $(1+\nu_m)/2$ the $\ell$ eigenvalues of $C_A$, the R\'enyi entropies can be expressed  as 
\begin{equation}
S_n(\ell)=\sum_{l=1}^\ell e_n(\nu_l)\,, \quad {\rm with }\quad
e_n(x)=\frac1{1-n}\ln \left[\left(\frac{1+x}2\right)^n+ \left(\frac{1-x}2\right)^n\right]\,.
\label{Sn}
\end{equation}
More details about this procedure can be found in e.g. Refs. \cite{vlrk-03,pe-09}. \footnote{
The above construction refers to the block entanglement in the fermionic degrees of freedom. 
However, in the case of a single block considered here, the non-locality of the Jordan-Wigner transformation 
does not change the eigenvalues of the  reduced density matrix because it mixes only spins within the block. 
This ceases to be the case when two or more disjoint intervals are considered \cite{atc-10,ip-10} 
and other techniques need to be employed \cite{fc-10} in order to recover CFT predictions \cite{fps-09,cct-09,gr-12,ctt-14}.  
}

The representation (\ref{Sn}) is particularly convenient for numerical computations: the eigenvalues $\nu_m$ of the $\ell\times\ell$
correlation matrix $C$ are determined by standard linear algebra methods and $S_n(\ell)$ is then computed using Eq. (\ref{Sn}). 
This procedure reduces the problem of computing the RDM from an exponential to a linear problem in the system size.
Also advanced analytic techniques are available to study the leading and subleading properties of the R\'enyi entropies
 \cite{jk-04,km-05,ccen-10,ce-10,fc-11,cmv-11}, but these will not be discussed here.

We are now interested in the relative entropies between the reduced density matrices of two different eigenstates.
Generically, these two reduced density matrices do not commute and so they cannot be simultaneously diagonalized to 
calculate the relative entropies from their eigenvalues in a common base. 
It is instead possible to use the composition properties of Gaussian density matrices \cite{fc-10}, i.e. of the form \eqref{gau},
to compute the traces of arbitrary products of these matrices. 
The technical details of this method are reported for completeness in the Appendix 
while in the following we limit ourselves to apply it to the cases of our interest. 
In this way, we can use free fermionic techniques to test the CFT predictions for the quantities 
$G^{(n)}(\rho_1 \| \rho_1)$, cf. \eqref{2ndreplicaSrel}, or equivalently  the R\'enyi relative entropies of integer order $n\geq2$. 
Consequently, they represent a very robust test on the validity of all the derivation presented in the previous section.


 \begin{figure}
 \centering
 \subfigure
   {\includegraphics[width=0.45\textwidth]{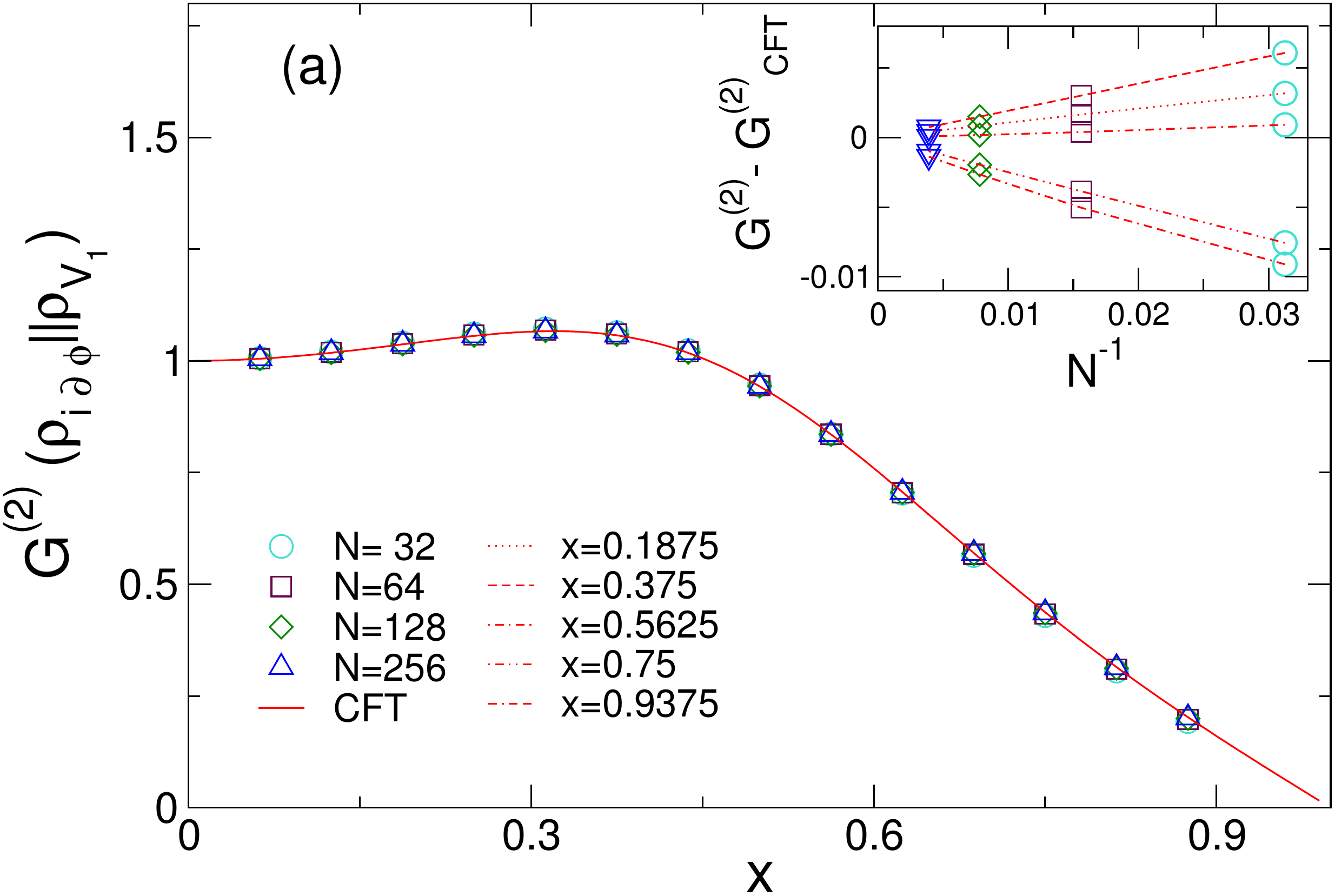}}
 \subfigure
   {\includegraphics[width=0.45\textwidth]{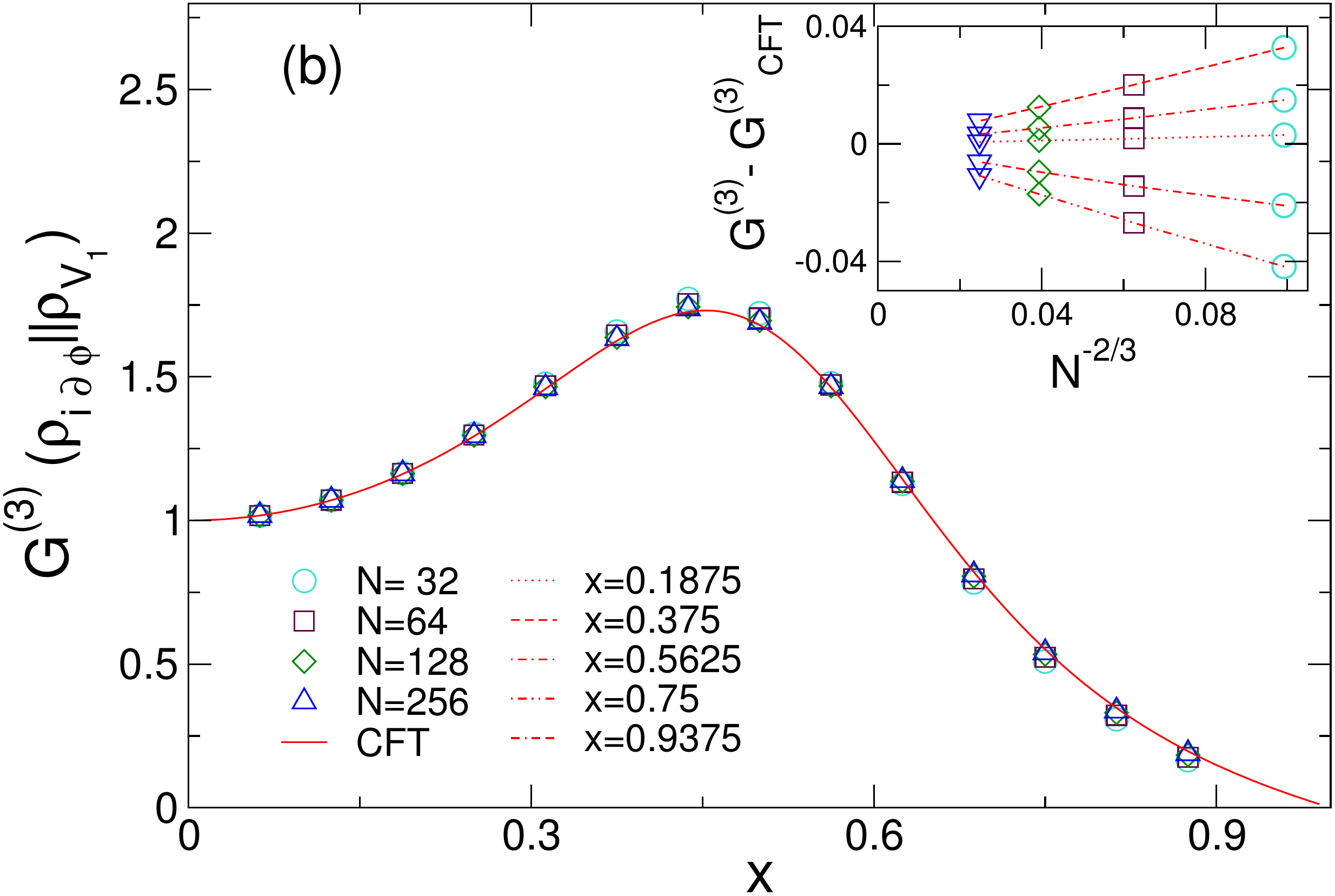}}
 \subfigure
    {\includegraphics[width=0.45\textwidth]{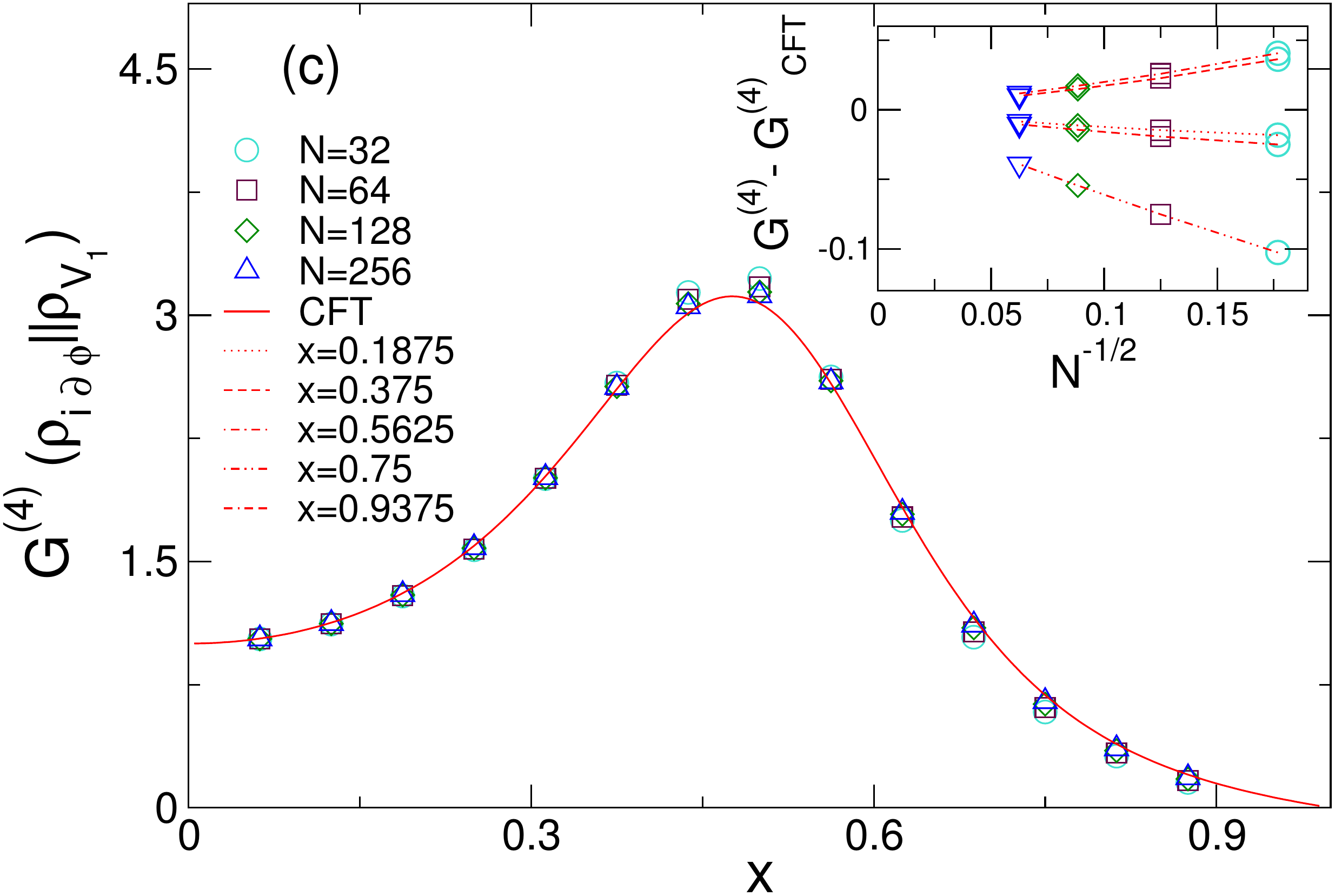}}
 \subfigure
   {\includegraphics[width=0.45\textwidth]{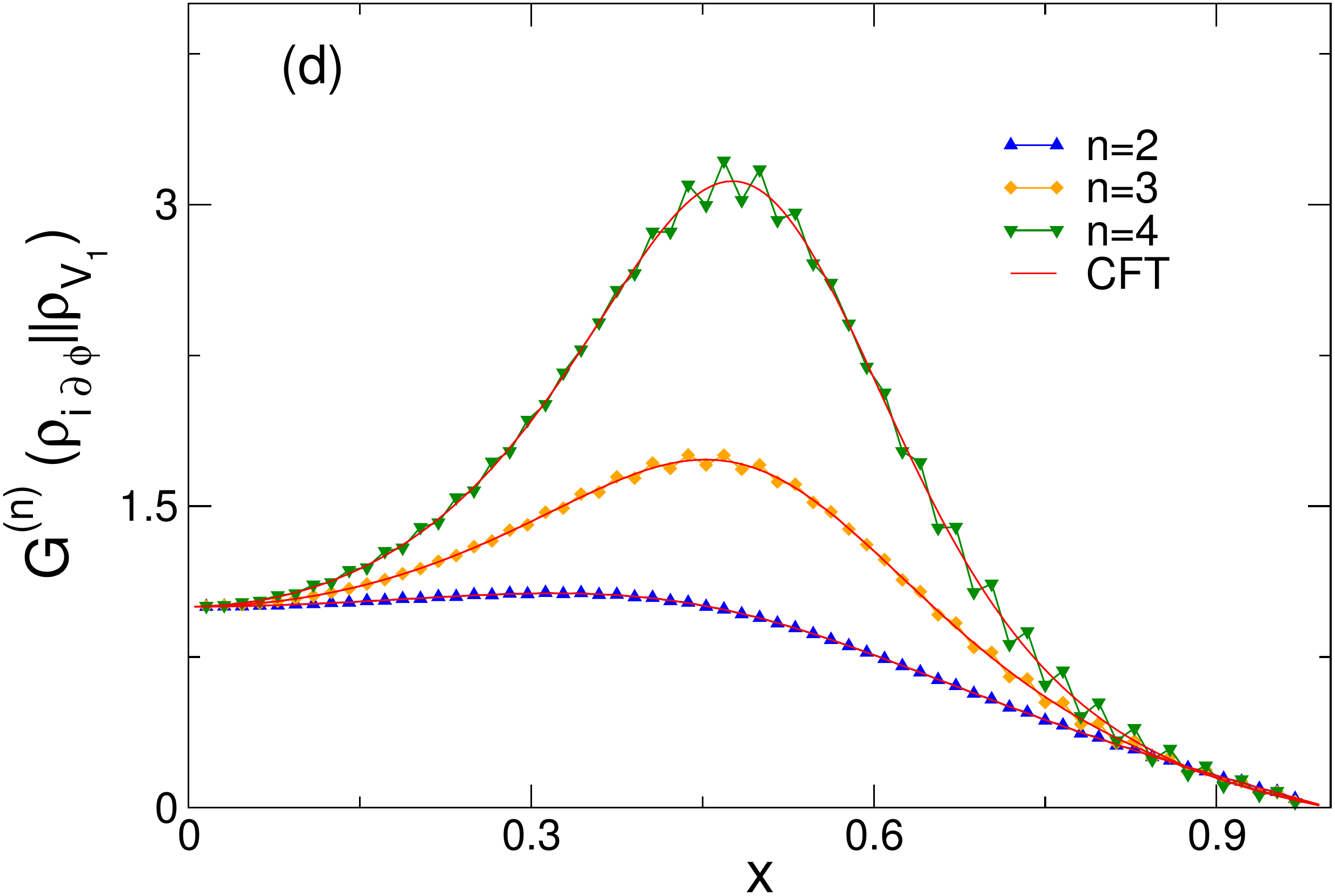}}   
 \caption{The quantity $G^{(n)} ( \rho_{i \partial \phi} \| \rho_{V_{1}} )$ as a function of $ x= \ell/N$ for different values of $n$. 
  The description of the Figure is the same as in Fig. \ref{Gn_VGS}.}
 \label{Gn_phiV}
 \end{figure}

\subsection{Numerical results}

With the techniques explained above and in the appendix we numerically compute the ratio 
\begin{equation}
\frac{{\rm Tr} \left(  \rho_1 \rho_0^{n-1} \right) }{{\rm Tr} \left( \rho_1 ^n \right)} 
\end{equation}
that in the limit $N\to\infty$ with $x=\ell/N$ kept constant should converge to the CFT predictions for $G^{(n)}(\rho_1\|\rho_0)$
in Eq. \eqref{2ndreplicaSrel}.
We consider the reduced density matrices $\rho_{1,0}$ corresponding to all the states for which we calculated the CFT 
predictions using the identification between lattice and CFT eigenstates in Eqs. \eqref{s1} and \eqref{s2}.

The numerical data for these $G^{(n)}$ between the chiral vertex operator $V_1$ and the ground state 
are shown in Figure \ref{Gn_VGS} for different values of $n$ and different system sizes. 
In the same figure we also report the CFT prediction $G^{(n)} ( \rho_{V_1} \| \rho_{GS} )$ 
(which we recall equals $G^{(n)} (\rho_{GS}  \| \rho_{V_1} )$).
It is clear also to the naked eye that the data converge to the CFT predictions by increasing the system size, but 
with a slower convergence for higher value of $n$. 

It is very interesting to study quantitatively the convergence of the data to the CFT prediction when increasing $N$
as shown in the insets of the figure for various $n$.
For the ground-state R\'enyi entropies $S_{n} (\ell)$  of free fermionic models, this convergence has been studied 
analytically in several works \cite{ccen-10,ce-10,fc-11} and it has been found to be of the form $N^{-2/n}$.
These corrections to the scaling found a CFT interpretation in Ref. \cite{cc-10} where it was understood that they originate 
from the local insertion of a relevant operator at the conical singularities defining the Riemann surface 
(alternatively can be thought as effects of the entangling surface \cite{ot-15,ct-16}).
Generically, in an infinite system they scale as $\ell^{-2\Delta/n}$ where $\Delta$ is the scaling dimension of the operator at 
the conical singularity and $\ell$ being the subsystem size. In finite systems, at fixed $x=\ell/N$, one can just replace $\ell$ by $N$.
For the XX model one finds $\Delta=1$ \cite{cc-10,ccen-10,ce-10}.
The same corrections of the form $N^{-2/n}$ have also been found for excited states \cite{sierra2012}. 
This is simply explained by the fact that the conical singularities are independent from the state,  
as studied in more details in \cite{c-16}.

Also in our study of the relative entropies, or more precisely of the quantities $G^{(n)}(\rho_1\|\rho_0)$, the structure of the 
Riemann surface is not altered by the presence of different fields generating the states.  
Thus one can safely conjecture that the leading  corrections to the scaling must be once again of the form $N^{-2/n}$. 
For $G^{(n)}(\rho_{V_1}\|\rho_{GS})$, this is confirmed to a great level of accuracy by the insets of Figure \ref{Gn_VGS}.

In the last panel of Figure \ref{Gn_VGS} we also study the effect of the parity of the subsystem size $\ell$. 
In the ground state (as well as in excited states), it is well known that the leading corrections to the scaling 
are not smooth functions of $x=\ell/L$ but they behave as \cite{ccen-10}
\begin{equation}
S_{n} (\ell) - S_{n}^{CFT} (\ell) \propto N^{-2/N} f_n(\ell/N) \cos (2 k_F \ell) .
\end{equation}
This oscillating term reduces to $(-1)^{\ell}$ at half filling $k_F= \pi/2$ (i.e. for zero magnetic field).
Again similar oscillations are expected  also for $G^{(n)}(\rho_1\|\rho_0)$, as 
confirmed by the data in Figure \ref{Gn_VGS} (d).

In Figures \ref{Gn_phiGS} and \ref{Gn_GSphi} we report the data for the replicated relative entropies 
between the ground state and the particle-hole excitation \eqref{s2} (corresponding in the continuum limit 
to the state generated by ${i \partial \phi}$). 
The overall discussion is very similar to the one above for the state generated by $V_\beta$ with the 
data approaching the CFT predictions $G^{(n)} ( \rho_{i \partial \phi} \| \rho_{GS} )$ and 
$G^{(n)} (  \rho_{GS} \|  \rho_{i \partial \phi} )$ for large $N$ as $N^{-2/n}$ (and also with pronounced parity effects 
in panels (d)). 
As in CFT, these functions are not symmetric under the exchange of the states in the relative entropy and in fact there 
is also a pronounced qualitative difference (already observed in CFT): 
while $G^{(n)} (\rho_{GS} \|  \rho_{i \partial \phi} )$ is a monotonous function of $x$ (as $G^{(n)} (\rho_{GS} \|  \rho_{V_1})$,
$G^{(n)} ( \rho_{i \partial \phi} \| \rho_{GS} )$ grows as $x$ increases from zero, has a maximum at a value depending on $n$
and then decreases. 

Finally in Figure \ref{Gn_phiV} we report the data corresponding to $G^{(n)} (\rho_{i \partial \phi}  \| \rho_{V_1})$.
Once again the data approaches the CFT perditions as $N^{-2/n}$ and we find that this is not a monotonous function of 
$x$, in analogy to $G^{(n)} (\rho_{i \partial \phi} \| \rho_{GS} )$.



\section{Discussion and future perspectives}
\label{section:conclusions}

In this work we applied the replica method \cite{lashkari2016} to work out several explicit examples of relative entropies between 
primary fields  of the free bosonic CFT, as well as the R\'enyi relative entropies \eqref{RRE}. 
The CFT results have been carefully tested against exact lattice calculations for the XX spin-chain, finding perfect agreement, 
once corrections to the scaling are properly taken into account. 
We must mention that we did not manage to work out the analytic continuation in the replica index for all the states we considered. 
Anyway, it is well known that finding the analytic continuation is not always an easy task and in some cases it is useful to resort to 
some approximations as e.g. \cite{dct-15}.
For the relative entropy, a possible approximation is the expansion for small subsystem presented in Ref. \cite{ugajin2016} which has 
also been extended to the case of disjoint intervals \cite{ugajin2016-2}, but its regime of validity is relatively small. 
A similar problem occurs also for the R\'enyi entropies of two disjoint intervals \cite{cct-09}.
In that case, among the many proposed approximations, an ingenious conformal block expansion 
has been considered \cite{gr-12} which turned out to describe effectively numerical data although the expansion is 
not systematic.  
It would be interesting to investigate whether some similar approach could be used also for the relative entropy. 

There are several generalization to our paper which could be worth investigating. 
For example, one  can consider other CFTs (such as Ising or other minimal models) as well as one can study different 
lattice models both free and interacting.
One could also deal with more general excitations, but when more complicated operators correspond to the excited states, 
the explicit calculations become very involved. 
For standard R\'enyi entropies, the extension of this kind of analysis  to 
descendants operators is reported in Ref. \cite{p-14} and in principle may be applied also to the relative entropy.
However, the calculation appears to be very cumbersome and difficult to extend to arbitrary values of the replica index $n$.


\appendix

\section{Correlation matrices of excited states in the XX spin chain and their product rules}
\label{appendixA}

In this appendix we report the generic correlation matrix of an eigenstate of the XX spin chain specified by a 
set of occupied momenta $K=\{k_i\}$ (see also \cite{afc-09,sierra2012}).
In order to make easy contact with Ref. \cite{fc-11} where the product rules of reduced density matrices have been reported, 
we work with the $2N$ spatial Majorana modes defined as
\begin{equation}
\begin{cases}
a_{2m -1} = c_m^{\dagger} + c_m, \\
a_{2m} = i (c_m^{\dagger} - c_m).
\end{cases}
\end{equation}
The Majorana correlation matrix $\Gamma^{K}_{m n}$  is defined as
\begin{equation} \label{Gamma}
\Gamma^K_{mn} = \langle a_m a_n \rangle_{K} - \delta_{mn},
\end{equation}
being $\langle \cdots \rangle_K$ the expectation value on the state labelled by $K$ 
(this matrix is trivially related to $C$ in the main text).
In matrix form it can be written as 
\begin{equation}
\Gamma^K_{m,n} = 
 \begin{pmatrix}
  \Pi_0 & \Pi_1 & \cdots & \Pi_{N-1} \\
  \Pi_{-1} & \Pi_0 & \cdots &  \\
  \vdots  & \vdots  & \ddots & \vdots  \\
  \Pi_{-(N-1)} & \cdots &  & \Pi_0 
 \end{pmatrix},
 \qquad 
 \Pi_m= 
 \begin{pmatrix}
  g_m^{(1)} & g_m^{(2)} \\
  -g_{-m}^{(2)} & g_m^{(1)}  
 \end{pmatrix}.
\end{equation}
In particular one has
\begin{equation}
\begin{cases}
g_{m-n}^{(1)} = \langle a_{2m} a_{2n} \rangle_K - \delta_{mn} = \langle a_{2m-1} a_{2n-1} \rangle - \delta_{mn} ,\\
g_{m-n}^{(2)} = \langle a_{2m-1} a_{2n} \rangle_K.
\end{cases}
\end{equation}

In order to evaluate these quantities one expresses the Majorana variables $a_n$ in terms of the fermionic ones $c_m$, 
whose correlation function are evaluated via Fourier transform, using the (trivial) correlation functions of the free fermionic variables.

By direct computation one finds
\begin{equation}
\begin{cases}
g_{m-n}^{(1)}= \frac{1}{N} \left[ \sum_{k \in K} e^{- i \frac{\pi k}{N} (m-n) }  + \sum_{k \notin K} e^{ i \frac{\pi k}{N} (m-n) }  \right] - \delta_{mn} \\
g_{m-n}^{(2)} =  \frac{i}{N}  \left[ -\sum_{k \in K} e^{- i \frac{\pi k}{N} (m-n)}  + \sum_{k \notin K} e^{ i \frac{\pi k}{N} (m-n)} \right]. 
\end{cases}
\end{equation}

\subsection{Product of reduced density matrices}

The algebra of Gaussian reduced density matrices is analyzed in Ref. \cite{fc-10}. 
In particular, it has been derived a \emph{product rule} to express the product of Majorana RDMs ($\rho_{\Gamma}$'s) in terms of operations 
on the respective correlation matrices ($\Gamma$'s).
If we implicitly define the matrix operation $\Gamma \times \Gamma'$ by
\begin{equation}
\rho_{\Gamma} \rho_{\Gamma'}= {\rm Tr} \left[ \rho_{\Gamma} \rho_{\Gamma'}   \right]  \; \rho_{\Gamma \times \Gamma'},
\end{equation}
then the following identity holds \cite{fc-10}
\begin{equation}
\Gamma \times \Gamma' = 1- (1- \Gamma')\frac{ 1 }{1 + \Gamma \Gamma'} (1- \Gamma),
\end{equation}
relating the correlation matrices of two RDMs to the one associated to their product.

Then the trace of two fermionic operators can be computed as (singular cases and ambiguities are discussed in \cite{fc-10})
\begin{eqnarray}
 \{ \Gamma, \Gamma' \} \equiv {\rm Tr} \left( \rho_{\Gamma} \rho_{\Gamma'} \right) = \prod_{\mu \in \text{Spectrum}[ \Gamma \Gamma']/2} \frac{1+ \mu}{2}.
\end{eqnarray}

Now, by associativity, one can extend to more than two RDMs
\begin{equation}
\prod_{i=1}^n \rho_{\Gamma_i} = \{  \Gamma_1 , \cdots , \Gamma_n  \} \rho_{\Gamma_1 \times \cdots  \times \Gamma_n},
\end{equation}
where
\begin{eqnarray} \label{iteration}
 \{ \Gamma_1, \cdots, \Gamma_n \} \equiv {\rm Tr} \left( \rho_{\Gamma_1} \cdots \rho_{\Gamma_n} \right)  = \{ \Gamma_1 , \Gamma_2 \} \{ \Gamma_1 \times \Gamma_2, \cdots  \}.
\end{eqnarray}
Eq. \eqref{iteration} can be used to iteratively evaluate traces of products of fermionic RDMs.




\end{document}